# THE EVOLVING PHOTOMETRIC LIGHTCURVE OF COMET 1P/HALLEY'S COMA DURING THE 1985/86 APPARITION

DAVID G. SCHLEICHER[1], ALLISON N. BAIR[1], SIOBHAN SACKEY[1,2], LORINDA A. ALCIATOR STINNETT[1,3], REBECCA M. E. WILLIAMS[1,4], AND BRIDGET R. SMITH–KONTER[1,5]

[1]Lowell Observatory, 1400 W. Mars Hill Rd., Flagstaff, AZ 86001; dgs@lowell.edu
[2]7725 N 109th Ave, Glendale, AZ 85307; [3]3507 Valley Haven, Kingwood, TX 77339
[4]Planetary Science Institute, 1700 E. Fort Lowell, Suite106, Tucson, AZ 85719
[5]Department of Geology & Geophysics, University of Hawaii, 1680 East-West Road, POST 813, Honolulu, HI 96822

ABSTRACT

We present new analyses of the photometric lightcurve of Comet 1P/Halley during its 1985/86 apparition. As part of a world-wide campaign coordinated by the International Halley Watch (IHW), narrowband photometry using standardized filters was obtained with telescopes at 18 observatories. Following submissions to and basic reductions by the Photometry and Polarimetry Network of the IHW, we further reduced the resulting fluxes to production rates and, following temporal binning, created composite lightcurves for each species. These were used to measure how the apparent rotational period (~7.35 day), along with its shape, evolved with time during the apparition. The lightcurve shape systematically varied from double-peaked to triple-peaked and back again every 8-9 weeks, due to Halley's non-principal axis (complex) rotation and the associated component periods. Unexpectedly, we found a phase shift of one-half cycle also took place during this interval, and therefore the actual beat frequency between the component periods is twice this interval or 16-18 weeks. Preliminary modeling suggests that a single source might produce the entire post-perihelion lightcurve variability and associated evolution, while an additional source probably also is required to explain additional features before perihelion. The detailed evolution of the apparent period varied in a non-smooth manner between 7.2 and 7.6 day, likely due to a combination of synodic effects and the interaction of solar illumination with isolated source regions on a body in complex rotation. The need to simultaneously reproduce each of these characteristics will provide very strong additional constraints on Halley's component periods associated with its complex rotation. To assist in these and future analyses, we created a synthetic lightcurve based directly on the measured data and how the lightcurve shape evolved week-to-week. This synthetic lightcurve was successfully compared to other data sets of Halley and provides a valuable estimate of Halley's activity even when no narrowband photometry measurements were obtained. We unexpectedly discovered a strong correlation of ion tail disconnection event start times with minima in the comet's gas production, implying that a decrease in outgassing is another cause of these events.

## 1. INTRODUCTION

Comet 1P/Halley is considered a special object for a number of reasons. Besides being the first comet recognized as reappearing on a periodic basis as well as the brightest of the periodic comets, it was also the first to be imaged from fly-by spacecraft (Sagdeev et al. 1986; Keller et al. 1986). Additionally, at least until the arrival of Hale-Bopp (1995 O1), it was the subject of the largest groundbased observational campaign in history. In spite of these efforts and numerous associated analyses, Halley has proven enigmatic and reluctant to give up its secrets, primarily because it was in a non-principal axis rotational state. Having never been seen before in a comet, it took several years before this "complex" rotation became accepted and even longer before many possible scenarios were narrowed down to a few that appeared to match the various constraints imposed by all of the data sets (cf. Belton 1990 and references therein).

With details of its rotational state uncertain, many anticipated findings for Halley also remained incomplete but attainable. Given this situation and the subsequent advancement in analysis and modeling capabilities, we have recently embarked on a new set of studies of Halley's coma and nucleus, with the intent of solving some of these long-standing mysteries. Our first new investigation, reported here, is that of Halley's rotational variability in outgassing and the evolution of the resulting lightcurve throughout its apparition in 1985/86. Subsequent studies will focus on the gas and dust jet morphology and on the creation of a nucleus model to reproduce these data sets.

Initial claims of a 2.2 day rotational period for Halley (Sekanina & Larson 1984, 1986) were followed by many other reports of a similar value before Millis & Schleicher (1986) discovered photometric evidence of a 7.4 day periodicity. This result was followed by other investigators finding the longer period in various types of data; a flurry of subsequent attempts to reconcile the two periods largely discounted the original 2.2 day period but found strong evidence for two component periods and the need for Halley to be in a non-principal axis rotational state (cf. Belton 1990; Belton et al. 1991; Samarasinha & A'Hearn 1991; and references therein). Our own initial effort to measure the rotational evolution combined narrowband data from four observatories, and yielded confirmation of the rapid evolution from a triple-peaked to a double-peaked lightcurve from 1986 March to April (Schleicher et al. 1990). It also revealed evidence of a decreasing effective period throughout the post-perihelion time frame (probably due to a synodic effect), but gave inconclusive results for the pre-perihelion interval due to the sparseness of the dataset.

To create the most complete homogeneous lightcurve possible, we have now incorporated all of the narrowband photometry submitted by observers around the world to the International Halley Watch (IHW) archive. This combined lightcurve of Halley's coma was then used to extract the apparent period as a function of time, investigate the evolution of the lightcurve shape, and ultimately produce a synthetic lightcurve that is our best estimate of Halley's behavior even when no observations were made. Additional constraints on the nature of Halley's rotational state are also presented.

## 2. THE DATA SETS AND REDUCTIONS

### *2.1 Methodology*

Because CCD cameras had only recently become available in the mid-1980s, had very small format size, and were difficult to operate in a routine manner, the vast majority of photometrically calibrated observations of Halley were obtained using the "tried and true" instrumentation of photoelectric photometers. A working group, chaired by M. A'Hearn, designed and had produced dozens of sets of narrowband filters, with filters isolating emission bands of OH, CN, $C_3$, and $C_2$, along with two ions and three continuum points. Known as the IHW filters, these sets were distributed to observers around the world. Following Halley's apparition, the resulting observations were submitted to the appropriate node of the IHW — in the case of narrowband photometry to the "Photometry and Polarimetry Network" (PPN) also headed by M. A'Hearn. All raw observations were reduced at this node in a common manner, resulting in absolute fluxes for the continuum, and for the emission bands following continuum subtraction. See A'Hearn (1991) for details of the entire process, Osborn et al. (1990) for information regarding the standard star calibrations, and A'Hearn & Carsenty (1992) or A'Hearn & Vanysek (2006) for the archived dataset.

We initially accessed the archive in 1992, but eventually identified problems with data originating from a few observatories as being inconsistent with observations from other sites. Most of these problems were resolved at the PPN node by 1995 when we again imported the database of reduced photometry; equivalent versions of the same data were moved to the Small Bodies Node of the Planetary Data System (PDS) where they remain available today. These fluxes form the starting point for the remainder of our analyses.

Because entrance aperture sizes varied widely for the photometers employed by the various observers, we could not just compare fluxes without applying some type of aperture correction. The most sensible method to do this was simply to calculate production rates for the neutral gas species and $Af\rho$ for the dust, since both quantities are aperture independent assuming the gas scalelengths reproduce the observed spatial distributions and the dust follows a canonical $1/\rho$ fall-off, where $\rho$ is the projected distance from the nucleus; no equivalent method exists for dealing with ions and so those data were excluded from further consideration. Each night's data from each observatory were further reduced using our standard procedures for applying $g$-factors, the Haser model with parent and daughter scalelengths, and daughter lifetimes for each neutral gas species, resulting in a gas production rate, $Q$, and the appropriate conversion factors to compute $A(\theta)f\rho$, a proxy for dust production (A'Hearn et al. 1984), for the continuum points (see A'Hearn et al. 1995 for details). Because Halley's phase angle, $\theta$, dropped to a very small value in 1985 November (<2°), we also applied a phase correction for the dust based on our derived phase function for Halley (Schleicher et al. 1998), yielding $A(0°)f\rho$.

We emphasize that $Q$ and $Af\rho$ are not true, instantaneous production rates but rather measures of the inner coma abundances of each species normalized for aperture size, although some trends with aperture size remain. Overall, Halley's total outgassing through much of the apparition was sufficiently high to result in greater outflow velocities than usual, and our standard scalelengths do not take this into account. This yields a slight upward trend in $Q$ for some species with increasing aperture size. Conversely, $A(0°)f\rho$ exhibited a downward trend with aperture size, especially before perihelion, because the dust in Halley had a steeper radial profile than $1/\rho$ — similar to many other comets. Because of the rotational variability, however, the radial distributions in the coma *never* follow the static equilibrium profiles, and larger apertures exhibit progressively longer phase lags and smaller amplitudes because a greater portion of the material (gas or dust) was emitted further back in time. This results in small apertures having higher apparent production rates than large apertures when outgassing increases after sunrise, but then reversing after outgassing declines near or after sunset. Because the aperture trends were constantly changing with time, we could not feasibly make suitable corrections. Fortunately, these issues are generally small (about 20% over the range of aperture sizes at a given time) as compared to the overall rotational variability but do add considerable scatter when examining the lightcurve in detail. We will discuss, as needed, the cases were this affects our results.

### *2.2 The Data Set*

In all, 2389 sets of narrowband photometry from 22 telescopes (located at a total of 18 observatories) were in the database we extracted from the IHW archive. With a policy to archive all data submitted to the IHW except for extreme outliers, it was inevitable that this would lead to the inclusion of “bad data” caused by non-photometric conditions, high airmass, mis-centering of the comet, etc. The detection of erroneous data points is made more difficult precisely because of the aperture trends noted above. To reduce the problems associated with a wide range of aperture sizes – 9 to 157 arcsec – we removed measurements obtained with aperture diameters smaller than ~25 arcsec, as these displayed the greatest trends with size and often had larger apertures measured near-simultaneously. We could not, however, remove the largest aperture data without leaving significant gaps in the lightcurve, and so these remained giving a total of 1938 sets.

While the filters used and the reduction methodologies were largely identical (cf. A'Hearn 1991), a very large variety of observing strategies were employed among observers. For instance, some observers used a high cadence of only a few filters, while others used a fixed, single aperture (some quite small and others very large), and still others (such as ourselves) used a range of aperture sizes. These decisions both greatly affect the resulting S/N of each data point and the amount of resulting scatter, even from a single site due to the specific apertures used. Despite the inherent scatter this created, by plotting all of the data we were able to identify 73 sets of data (including all 27 sets from a single site) as being clearly discrepant, and these were removed from further consideration, yielding 1865 sets from 17 sites — the locations and associated statistics are given in Table 1, while a complete list of observers is contained in Appendix C of A’Hearn (1991). Unfortunately, there were several other data points that we were suspicious of but could not definitively be identified as “bad,” either because no other data existed for comparison, or two contemporaneous points clearly disagreed with each other but we couldn’t determine which was the bad one. Rather than remove all of these, we chose to include them and remain aware of their situation. Note that all but one of these cases were prior to perihelion when less data were obtained, lightcurve amplitudes were smaller, and Halley’s brightness and the resulting S/N was lower.

[TABLE 1 HERE]

Because Halley's rotational variability was relatively slow, we next averaged all data from a particular site into ~2 hr bins. This served to treat more equally measurements obtained at very different cadences (each binned point contains between 1 and 36 individual points), and averaged out aperture effects for those sites where multiple apertures were employed. A final total of 461 binned sets resulted from this operation. The diversity in how data were obtained, combined with the real dispersions associated with aperture effects, make it difficult to assign appropriate uncertainties to be associated with each binned set. In fact, photometric uncertainties are generally negligible compared to aperture trends. Therefore we have not propagated sigmas nor plotted errorbars but simply note that when significantly different aperture sizes are involved, considerable dispersion is also likely to be present.

The final steps prior to lightcurve analyses began with the removal of the secular trends during the apparition. While there are good reasons to expect a comet to follow a near-linear relation in log $Q$ vs log $r_H$, where $r_H$ is the heliocentric distance, we also tested fits in log $Q$ vs time from perihelion, $\Delta T$; however, these were not as good as those using log $r_H$ and were discarded. Although first order fits of log $Q$ vs log $r_H$ were adequate for some species, other species were clearly better fit in second order, and so in all cases the second order solutions (see Table 2) were used to compute the resulting residuals, $\Delta$ log $Q$. Plots prior to secular trend removal of our own, more limited data set were presented by Schleicher et al. (1998), and so such plots are not included here; these confirm the well-known fact that Halley is more productive following perihelion than prior over a wide-range of heliocentric distances.

[TABLE 2 HERE; INTENDED TO FIT IN A SINGLE COLUMN]

We next selected which species were most useful both for period determinations and for creating our best overall lightcurves. As originally noted by Millis & Schleicher (1986), all measured species showed the same basic rotational variability, but with differing amplitudes and phase lags due to differing lifetimes (among the gas species) and velocities (dust is much lower than any of the gas species). Although OH has by far the highest production rates, measurements of this species were very undersampled since observers at many sites did not observe near the atmospheric cut-off. The red continuum was also not observed from many locations because most phototubes had little or no sensitivity in the red, while the UV continuum was both undersampled and had much lower S/N than the green continuum. As a result, analyses were performed using three gas species — CN, $C_2$, and $C_3$ — along with the green continuum (4845 Å), and the values for these are listed in Table 3.

[TABLE 3 HERE (TWO COLUMNS PER PAGE, 4 PAGES TOTAL)]

## 3. LIGHTCURVE ANALYSES

### *3.1 Characteristics of Halley's Lightcurves*

The resulting $\Delta$ log $Q$s and $\Delta$ log $A(0°)f\rho$s are plotted as a function of time in the four panels of Figure 1. Immediately evident are the natural gaps in the data, due to solar conjunction surrounding perihelion and the avoidance of the full moon by most observers each lunation. The reasonableness of the secular trend removal is evident from these panels as there are no significant trends either near or far from perihelion and each species shows similar distributions of the residuals about zero. Several other characteristics are also evident. In particular, the amplitude of the envelope is greatest in April (+50 day $< \Delta T <$ +70 day) coinciding with the comet's closest approach to Earth; this is simply because the projected aperture sizes (color-coded in the figure) were smallest at this time, resulting in the least dilution from older material. Even with similar projected apertures, however, the measured amplitudes are generally larger after perihelion than prior and we suspect that this is associated with the generally higher production rates post-perihelion; if one or more source regions on the nucleus receive more hours of illumination from the Sun and/or have the Sun closer to the zenith, then higher $Q$s and a larger amplitude lightcurve could result.

[FIGURE 1 HERE]

Finally, $C_3$ consistently exhibits the largest amplitude, then CN, $C_2$, and dust, because of progressively longer parent lifetimes or, in the case of dust, lower outflow velocities; the very strong aperture trend in $\Delta$ log $A(0°)f\rho$ values before perihelion is readily apparent, making meaningful analysis of the combined dust lightcurve nearly hopeless.

*3.2 The Evolving Nature of Halley's Lightcurves*

To better understand the detailed evolution of Halley's lightcurve(s), we must examine a greatly expanded view such as we provide for $C_3$ in Figure 2. Here, each 4-week interval of the apparition is plotted in a separate panel. Projected aperture sizes are again distinguished by color. A synthetic curve, based on basic properties of the observed lightcurve and discussed in the next sub-section, has been overlaid. Immediately evident are some of the characteristics already noted such as the larger amplitudes early post-perihelion and differences associated with aperture size. Also evident are the changes in the lightcurve shape, particularly from a triple-peak in early March (+21 to +35 day), to a double-peak in early April (+46 to + 63 day), and back to a generally triple-peak appearance in late-April and early-May (+70 to +91 day) and a double-peak in late-May and early-June (+91 to +112 day). The situation is more difficult to interpret prior to perihelion due to more sporadic coverage along with the generally smaller amplitudes, but the overall pattern of a slow oscillation between double and triple-peaked variability seen after perihelion is also present before.

[FIGURE 2 HERE]

We, therefore, both confirm and extend the findings by Schleicher et al. (1990) that the lightcurve shape generally varied in a slow and systematic manner. However, the oscillatory nature of the variations is inconsistent with the explanation for this behavior proposed by Schleicher et al. as being due to a seasonal effect associated with the changing orientation of the Sun. In that scenario, changes should occur more slowly when the comet is further from the Sun and its angular motion with respect to the Sun decreases. What we now identify, instead, is that each 7+ day "rotational" cycle varies its overall pattern from double, to triple, and back to double-peaked shape every 8-9 weeks throughout the apparition. We directly attribute this behavior to Halley's non-principal axis rotational state, with the observed oscillation period associated with the beat frequency between the two component periods. We return to this in more detail in later sub-sections.

The basic observed, i.e. apparent, 7+ day periodicity is also presumably due to a combination of the component periods along with the effects due to the changing orientation of the Sun, and we will investigate this further in Section 3.5. The relevant issue here is that the measured periodicity also varied throughout the apparition, but within a relatively small range of 7.21 to 7.59 days, and the bulk of this variation is presumably a synodic effect that is greatest near perihelion. Thus both the shape and periodicity of the lightcurve varied relatively slowly, properties we utilize when creating a synthetic lightcurve as described next.

*3.3 The Creation of Synthetic Lightcurves for Halley*

Although there are many time intervals during Halley's apparition when its photometric lightcurve is well characterized, there are numerous other intervals where little or no data were obtained (see Figure 2) despite a world-wide observational campaign. We, therefore, pursued the creation of a synthetic lightcurve to fill-in the gaps in coverage. Because the lightcurve shape generally varied little from one week to the next, we realized that surrounding weeks could be used to fill in gaps for a given "rotational" cycle, allowing for the evolutionary trends at each "rotational" phase. Using the measured periodicity, we phased the data and then offset each cycle to obtain a stack of curves. Ideally, we would have next fit this with a 2-dimensional rubber sheet, with differing amounts of flexibility in the phase and cycle number dimensions, but we did not have access to an equivalent algorithm and so we instead used the workaround described next.

Our kludge begins by applying a smoothing spline to the unphased data, yielding a smoothed value at uniform intervals of ~2 hr, comparable to the original binning. Each of these new values were next assigned a weight based on their respective distance (in time) from the real data; a value of unity was assigned if the change in $\Delta T$ was less than 2.4 hr or if multiple points were measured at slightly larger offset values of $\Delta T$, with decreasing weights set until reaching a weight of zero if the nearest points had a change in $\Delta T > 16$ hr where we conclude that the smooth spline is non-constraining for such a large extrapolation. The complete curve (along with the assigned weights) was next phased according to our mean period (see Section 3.5), yielding a phase value and cycle number for each point of the curve. We then fit a series of weighted smooth splines in the cycle dimension, with one fit at each phase value for intervals of 0.005 phase. Note that this step provides the interpolations from surrounding cycles when observations were missing at a particular phase for a given cycle. These new values replace the original smooth spline for each cycle, and these curves are "unwrapped" to create a single new curve. As this new curve has considerable jitter caused by each phase fit having been created independently of its neighboring phase values, we finish by re-smoothing the complete curve.

While not ideal, this entire process mimics applying a two-dimensional smoothing spline with different amounts of smoothing in the phase and cycle dimensions. It works primarily because the evolution of the lightcurve is relatively slow compared to the basic periodicity. Even the change in the basic periodicity due to the synodic effect is effectively treated as an evolution of the lightcurve, since the maximum net offset in phase from one cycle to the next due to using the mean value of 7.35 day rather than the specific period for a given cycle was <0.03, while the accumulated maximum offset was ~0.07 phase. Several practical issues were addressed by trying a wide range of smoothing values at each step of the process, constantly comparing the results with the actual data and iterating. Our final amounts of smoothing varied among species due to differences in amplitude and S/N. Due to the large gap in data during solar conjunction, we fit the pre- and post-perihelion data separately. There remain a few brief intervals in the lightcurve where our final synthetic lightcurves do not agree with our "by eye" evaluation of a proper interpolation between cycles, but we have chosen to *not* manually override the synthetic curve in these cases, and instead simply note them in the next sub-section. Additionally, there are some intervals, especially pre-perihelion, where discrepancies among the data make the final curve difficult to assess.

Finally, as a check of our results in some of the sparse data intervals, we compared two non-narrowband data sets to our synthetic curves. The first was V-band photometry obtained on 26 nights during a 40-day interval in March/April by Neckel & Münch (1987). These large aperture (317 arcsec) measurements are dominated by a combination of $C_2$ emission and continuum, and as expected exhibit a smaller amplitude and increased phase lag over this paper's smaller aperture data, but are otherwise in complete agreement with our synthetic curves. The second data set, conversely, consists of measurements using a very small aperture (18 arcsec square) obtained with the International Ultraviolet Explorer (IUE) satellite by Feldman et al. (1987). The Fine Error Sensor (FES) camera on IUE had a bandpass of about 4000-6500 Å and so included $C_3$ emission in addition to $C_2$ and continuum. The very small aperture resulted in larger amplitudes and smaller phase lags than our own data that, when allowed for, again matched the synthetic curve very well including intervals near full moon where very little groundbased data were obtained. Even a brief, small apparent outburst detected late on March 18 (McFadden et al. 1987) appears to simply be the forerunner of a small developing shoulder in the lightcurve at phase 0.10 on subsequent cycles (+8 and +9). Overall we, therefore, consider the quality of the synthetic curves to be extremely good.

### *3.4 The Detailed Evolution of the Shape of Halley's Lightcurves*

The creation of synthetic lightcurves as just described greatly assists our more detailed investigation of Halley's evolving lightcurve shape simply because most of the gaps in observational coverage are filled in with appropriate interpolations. In Figure 3, we show the phased results for $C_3$ again using the average apparent periodicity of 7.35 days, and shifting each successive cycle downward. All of the previously identified characteristics are either clearly (amplitudes, slow evolution) or more subtly evident (double/triple peaked oscillation). For instance, looking first at the post-perihelion curves, one sees a strong minimum near phase 0.75 followed by a strong maximum near phase 0.90 at cycle 4 and persisting until cycle +11 when both features begin to disappear and are gone by cycle +13, with a minimum replacing the maximum by cycle +15. A second maximum, near phase 0.50, appears to persist throughout the interval, though with some changes in shape, while the third peak (at phase 0.30) in cycle +4 rapidly becomes a minimum only 5 cycles later. Similar behavior is seen before perihelion, where a maximum near phase 0.35 on cycle -12 evolves into a strong minimum by cycle -5, a maximum near phase 0.6 appears to persist until Halley disappears behind the Sun, and a valley near 0.15 becomes a peak.

[FIGURE 3 HERE (2 FULL-PAGE, FACING PANELS; FIGURE CAPTION GOES WITH PANEL A)]

The previously mentioned double/triple peaked oscillation is most obvious in the larger amplitude data after perihelion, with a triple-peaked lightcurve in cycles +4 and +5, and again between cycles +10 and +12, while the double-peaked curve is evident between these times and again at the end of the apparition. Pre-perihelion data exhibit a triple peak curve near cycle -12 and by cycle -5, while there is a double-peak appearance at cycles -8 and -7; however, the resulting curve is distorted by additional minor features that may or may not be real. (Note that some artifacts of the process to create the synthetic curve are obvious, such at the very sharp rise at phase 0.75 on cycles -11 and -10, or the small bump at phase 0.32 on cycle -8, and the unusually shallow dip at phase 0.75 on cycle +7 when surrounding cycles exhibit a strong minimum.)

Another repeating characteristic is that the spacing between features is never equal; double-peaked features are typically separated by ~0.4 phase (or 0.6 phase in the other direction), while triple-peaked cases usually have one pair separated by only 0.20-0.25 phase. Fortunately, this behavior greatly assists in making comparisons of the lightcurve from different epochs, and we find multiple instances where the gross shape is repeated after 7-9 cycles, but with an unexpected phase shift of about 0.6 (or 0.4) cycle. Thus the two peaks seen in cycles +7 through +9 at 0.55 and 0.90 phase reappear in cycle +16 at phases 0.15 and 0.50, while the multiple features seen in cycle +5 are also evident in cycle +12 with a ~0.6 phase shift. (Errors in the apparent periods were readily eliminated as a possible cause of the phase shift since they would have needed to average 0.4-0.5 day *per cycle* to accumulate to a half-cycle shift after only 7-9 cycles.) We will return to this phenomenon after examining how the apparent period itself varies during the apparition.

### *3.5 The Detailed Evolution of Halley's Apparent Period*

As just seen, the shape of Halley's lightcurve is constantly evolving, and this behavior makes our initial goal of precisely determining the apparent period and how it changes during the apparition more difficult than one might expect. There were already difficulties with applying standard period measurement techniques given both the changing amplitudes and phase lags of lightcurve features due to the large variations in the projected aperture sizes throughout the apparition. Long-term variations in the lightcurve characteristics, presumably due to seasonal effects, make even the expected repetition of the gross shape of the lightcurve after ~15-18 weeks (twice the half-phase-shift intervals described in the previous section) difficult to recognize. We also fully expected to see a clear signature of the synodic effect on the period as Halley approached and receded from the Sun. When taken in combination, there are too many effects that actually alter the lightcurve shape and period for search algorithms to properly interpret the periodicity over long intervals; Fourier techniques will find non-existent periodicities or component values rather than the apparent period, while phase dispersion minimization (PDM) techniques will be fooled by the lack of repeatable features. For these reasons, we have chosen to measure much shorter intervals over which many of these complicating effects are minimized, and where we can identify by eye which portions of the phased lightcurve are relatively stable and which portions are evolving and therefore non-constraining when determining the apparent period.

Ultimately, we measured the periodicity using 21 days of data at a time, a convenient unit encompassing nearly three complete 7+ day "rotational" cycles, with overlapping intervals shifting by one week each, thereby having each week of data used within three successive intervals. For each interval, we interactively phased the lightcurve – both the data and the synthetic curve – for a series of trial periods to find the value that gave the best match to the persistent peaks and troughs, and paying least attention to the portions of the lightcurve that were evolving. In each case, we also estimated a viable range of periods, and these were used to yield an estimated uncertainty. Period searches were performed for each species independently and, within the uncertainties, the different species gave consistent results (note that green continuum was not included for the pre-perihelion time frame for the reasons previously mentioned). The best values for each interval, along with the associated uncertainties, are listed in Table 4 and, for $C_3$, are shown in Figure 4.

[TABLE 4 HERE]

[FIGURE 4 HERE; INTENDED TO FIT INTO A SINGLE COLUMN]

As discussed previously, the post-perihelion data are more complete and the lightcurve is more definitive. For the time frame encompassing +14 to +126 day (having mid-points of 24.5 to 115.5 day), the best fit periods ranged from 7.21 to 7.57 day, and the mean value from all of the 3-week intervals ranged from 7.347 to 7.361 day among the species, for an overall average of 7.35 day that was used in the phase plots in earlier sub-sections. A clear decrease in the period is evident, and is consistent both with the similar trend initially reported by Schleicher et al. (1990) for the March/April time frame and with their suggestion that this was most likely associated with a synodic effect for an object in prograde rotation. In the context of complex motion, this in turn corresponds to the prograde direction for the nucleus's precession, rather than its rotation, but one still expects the peak in the apparent period to occur near perihelion when the comet's rate of change in its true anomaly is greatest.

An unexpected finding is the stair-step type decrease after perihelion, with two significant drops each followed by many weeks of near-constant values for the apparent periodicity, and we return to this unusual characteristic in the next sub-section. Turning to the time frame prior to perihelion, as previously seen the situation is muddled due to the generally smaller amplitudes and evidence for additional smaller features also changing with time. This directly resulted in more ambiguous period determinations and correspondingly much larger viable ranges and inferred uncertainties. Final resulting periods varied between 7.39 and 7.59 day for the time frame encompassing -105 to -28 day (having mid-points of -94.5 to -38.5 day), with an overall average of 7.49 day. In the month leading up to Halley's conjunction with the Sun and perihelion, the apparent period increased as one would expect in the prograde case but surprisingly the preceding month exhibited a downward trend (see Figure 4), and the resulting mean period, 7.49 day, is higher before than after perihelion. We also defer discussion of this unusual behavior to the next section.

As noted, the PDM methodology did not provide meaningful period determinations when used for long intervals of data, but we found it to be a useful check on the periods we obtained from visual inspection when the intervals were only 3 weeks in length. PDM calculates the overall variance in the phased data for each of a series of trial periods (see Schleicher et al. 1990 for examples). We used the same 3-week intervals as above, but even within this short interval the program often had difficulty finding a period solution. This is because it tries to match the repetition of the lightcurve shape, which is sometimes different from cycle-to-cycle. Even so, the PDM solutions, while giving a wider range for the best answer, are usually reasonable compared to those determined by visual inspection of the lightcurve for the post-perihelion data (Table 4). For the pre-perihelion interval, the viable ranges given by PDM do not match our visual results as strongly, and the results are more scattered with larger uncertainties. Overall, PDM gives misleading solutions when there are large data gaps, when the amplitude is small, and during transitions in the number of peaks when the lightcurve is quickly evolving.

### *3.6 Additional Constraints on Halley's Nuclear Properties*

We can now combine our results with others to determine further constraints on Halley's rotational state. At the end of Section 3.4 we noted that the lightcurve shape tended to repeat after 7-9 cycles but with a phase shift of about 0.6 (or 0.4 in the opposite direction). Taking into account the synodic effects just detailed, the accumulated phase shift with respect to using the mean period of 7.35 day is about 0.07. Adjusting for this effect implies that the pattern actually shifted by one-half of a cycle, a more sensible amount even though the specific cause for a half-cycle shift in Halley's lightcurve eluded us for quite some time. We will return to this topic below.

We previously noted our suspicion that the 3-to-2-to-3 peak evolution is associated with having the two component periods being somewhat offset from a 2:1 ratio; in fact, the only evolution that should exist if the component periods were exactly commensurate would be a much slower seasonal variation, so therefore an offset must indeed exist and is directly associated with the continuing evolution over the 9 months of the apparition. In the scenario favored both by Belton et al. (1991) and Samarasinha & A'Hearn (1991), the precession component has a period of ~3.7 day while the roll about the long axis has a very poorly constrained period of ~7.3 day; see Figure 1 of Belton et al. With this combination of component periods the nucleus would have rolled about the long axis once plus an additional ~5° in the 7.4 days needed to precess twice, progressively changing its amount of illumination and only returning to its starting position in both precession and roll after 18 months (for this calculation, we intentionally ignore synodic effects, though they also become important). It is therefore immediately obvious that the offset in commensurability of the component periods must be considerably larger than 0.1 day to produce the beat frequency we observe. Because the rate of evolution is directly correlated to the amount of the offset, we can utilize this frequency to determine the offset and, if one component period is known, compute the value of the other component.

In fact the orientation of Halley's long axis at the time of the Vega 1, Vega 2, and Giotto encounters provides the strongest constraint on the possible rotational states, the associated precession period, and the direction of the total angular momentum vector (see Belton 1990, Belton et al. 1991, Samarasinha & A'Hearn 1991), and these authors conclude that the precession period must have a value of near 3.7 day to match the orientations and to return the nucleus to approximately the same orientation with respect to the Sun after ~7.4 day. In particular, Samarasinha (personal communication) estimates the range of possible values from 3.5 to 3.9 day but most likely between 3.6 and 3.8 day (where an asymmetric rotator is assumed), while we infer an approximate uncertainty in Belton et al.'s (1991) 3.69 day period of ±0.03 day (where a symmetric rotator was assumed) given their claimed uncertainties in the determinations of the long-axis directions at each fly-by. We will return to this later but initially assume a value of ~3.7 day. In contrast, the rotation period around the long axis, i.e. the roll period, has no strong constraints other than that imposed by the lightcurve – the need for source regions on the nucleus to approximately return to the same orientation with respect to the Sun in 7+ day – and these authors each conclude that the roll period must be near 7.1-7.3 day or, less-likely, near 2.4-2.5 day (the remaining intermediate near-commensurate value of 3.7 day is ruled out as the lightcurve would show a 3.7 day rather than 7.4 day signature). Here we initially assume their preferred "slow" solution near 7.3 day, but as discussed in the prior paragraph, the non-commensurability offset in this case must be significantly greater than 0.1 day.

Initially we thought that this offset simply needed to be enough to progressively roll the nucleus an extra 360° in 8-9 weeks, the time it took to evolve from 3-to-2-to-3 peaks. But in this case the one-half cycle phase shift remained unexplained, and we then examined double this interval – 16-18 weeks – where two of these one-half cycle shifts would return the source to its starting orientation. Here an additional roll of ~22° every 7.4 day would return the nucleus to the starting position, thereby implying that the rotational period was ~6.95 day. The breakthrough came when it was realized that this new scenario also provides a natural explanation of the one-half cycle phase shift in 8-9 weeks: after 8-9 weeks, a source on the nucleus would have rolled an extra ~180°, but after *one* additional precession period of 3.7 day the source would *then* have the same orientation with respect to the Sun as when it started, yielding the one-half cycle shift! With the mystery explained, we conclude that orientations repeat after an even number of precession cycles every 16-18 weeks and that the rotation period is offset from twice the precession period by ~0.45 day. Note that thus far we have assumed that the rotation period is less than 2:1 of the precession period, i.e. ~6.95 day, but that a corresponding offset greater than 2:1 has not been ruled out, i.e. ~7.85 day.

The prior results are also intertwined with the synodic behavior discussed in Section 3.5. The changes in Halley's apparent period are generally consistent with a net pro-grade motion of the nucleus as it orbits the Sun, where the bulk "rotation" has the same sense as the orbital motion, and the time required to go from local noon to the next local noon increases as the orbital angular motion increases near perihelion. (Note that because we are observing material released into the coma from nucleus source regions, it is the length of the solar "day" that is relevant, and *not* the change in viewing orientation from the Earth.) The details of the functional form vary with the obliquity and direction of the rotation axis even for a body in simple rotation, but in the case of Halley's non-principal axis rotation the details can become quite messy; we first examine the easy case where a simple rotator has its pole in the direction of Belton et al.'s (1991) preferred total angular momentum vector, RA = 6° and Dec = -61°, corresponding to an obliquity of 17° and an angle about the orbit from perihelion of 338°. We find that the peak synodic or solar period occurs only 1 day prior to perihelion, because of the low obliquity coupled with an angle of the pole about the orbit only 22° from the Sun. We also find that the synodic period at the start and end of the apparition for which we have photometric measurements would only be 0.02 day longer than the nominal sidereal period. If one further assumes an effective sidereal period of 7.30 day, then the peak synodic period would have been 7.80 day. This example curve is overlaid to our extracted apparent periods in Figure 4, and it is evident that it provides a very reasonable first-order match to the measured periods.

A more detailed comparison, however, even more clearly reveals the other characteristics identified earlier – the bulk pre-/post-perihelion asymmetry and the stair-step changes in the apparent periods. In spite of the relatively large uncertainties on each pre-perihelion measurement, we think that the systematically higher values are both real and a natural consequence of the complex motion, even though we cannot currently provide an explanation. When looked at as a difference from the simply synodic curve, the stair-step behavior drops are actually followed by an increasing trend in the period over several cycles, and only look level after perihelion because the general synodic trend is downward and is thus canceled out. In spite of not having a ready explanation for the cause, we are certain that it is directly associated with the complex motion and particularly with the beat frequency based on a simple fact: the timing of each of the downwards steps — at -70, +42, and +91 day — coincide with each of the times when the lightcurve changed from triple to double-peaked; an additional step is expected to have taken place between about -10 to -20 day, but no observations exist to examine this possibility. While we can conjecture that the reduction from three to two peaks in the lightcurve is due to a source that had been illuminated at a particular precessional phase having progressively rolled further each cycle and is no longer illuminated at that phase, it is unclear why this should cause a downward step in the derived period, to be followed by a steady increase in the derived period for the next several cycles. As suggested by M.Belton (personal comm.), a contributing factor might also be the "nodding" motion, in addition to the rotational and precessional motions, expected for an asymmetric body in non-principal axis rotation. We can only assume that detailed modeling will reveal how these various pieces of the puzzle go together.

## 4. DISCUSSION AND SUMMARY

Considerable progress in understanding Halley's unique characteristics was made in the years immediately following its 1985/86 apparition and the first close-up imaging of a comet from fly-by spacecraft. In particular the earlier inference (Cruikshank et al. 1985) and discovery (Millis et al. 1985) that comet nuclei had much lower albedos and correspondingly larger sizes than had been assumed was dramatically confirmed by the spacecraft (Sagdeev et al. 1986; Keller et al. 1986), while the long-known brightness asymmetries about perihelion were confirmed with measurements of gas and dust production rates from Earth. Our discovery of the ~7.4 day periodicity in the lightcurve (Millis & Schleicher 1986) was followed by other evidence for this period in differing data sets, particularly coma morphology (cf. Samarasinha et al. 1986; Hoban et al. 1988). These data sets also established the need for the nucleus to return to nearly the same orientation each cycle for all rotational phases, while the slowly changing lightcurve shape implied some type of continuing evolution. Even brightness measurements extracted from the 1910 apparition showed evidence for a 7.4 day periodicity and a triple-peaked lightcurve (Schleicher & Bus 1991), implying that Halley's basic behavior is stable. We also knew that all of the gas species measured in the visible/near-UV, along with the dust, exhibited the same basic brightness variations, thereby implying their parents had a common origin; small differences in amplitude and slight phase shifts between species are entirely consistent with differing parent lifetimes and, in the case of dust grains, slower outflow velocities. Considerable other evidence led to the conclusion that Halley's nucleus is not in simple rotation, but instead is in a non-principal axis, i.e., complex, rotational state (cf. Belton 1990). Furthermore, no complex scenario having a 2.2-day component period can return the nucleus to the same orientation every 7+ day and are therefore excluded.

Our new work, utilizing all narrowband photometry measurements submitted to the IHW, has resulted in a number of discoveries presented in Section 3, but is somewhat limited by the diverse methodologies used among the observers around the world. In particular, due to the 10× range in aperture diameters employed, strong aperture effects dominate the dispersion in data even when taken near-simultaneously. While the large lightcurve amplitudes seen in March and April overwhelm the aperture effects, this aperture-caused dispersion makes it much more difficult to interpret the lightcurve evolution at other times when the overall amplitudes are smaller. Thus in hindsight, despite the use of matching narrowband filters, the heterogeneous nature of the data acquisition proved to be a significant limitation on what could be readily extracted from the combined dataset. The pre-perihelion time frame proved unexpectedly problematic – we had assumed that the difficulty in obtaining definitive period determinations by Schleicher et al. (1990) was primarily caused by an insufficient number of measurements, but we discovered that the lightcurve shape had significantly lower amplitudes and often more features per cycle than were seen after perihelion. These factors, coupled with generally lower S/N because the comet was fainter, resulted in a less clear understanding of the lightcurve evolution and correspondingly less certain values for the extracted apparent periods.

### *4.1 Evolutionary Effects*

Despite these problems, our new analyses using the entire IHW narrowband photometry data set indeed allowed us to greatly extend and improve our understanding of Halley's inner coma lightcurve over what was known. Notably our analyses have revealed that throughout the post-perihelion time frame the number of features oscillated from 3-to-2-to-3-to-2, while a similar oscillatory pattern is evident but less definitive before perihelion. The discovery of this oscillation was facilitated by the persistent non-equal spacing between different pairs of lightcurve features – not only did the number of features oscillate, but their relative spacing repeated at the same stage of the oscillation. Moreover, while the overall pattern repeated after about 8-9 weeks – an unexpectedly rapid rate given proposed component periods – it also shifted by about one-half of a 7+ day cycle. We, therefore, surmise that the true repetition period is twice this, i.e. about 16-18 weeks, and that this is directly associated with the degree of non-commensurability between the complex component periods. Indeed, if the component periods were commensurate then no evolution other than seasonal changes would exist.

Unfortunately, this evolution of the lightcurve shape makes the extraction of the apparent period as a function of time considerably more difficult as one needs to choose which features of the lightcurve one should align when phasing the data. We ultimately chose to match those features that were evolving most slowly (for a particular time interval), and largely ignore more rapidly varying features. Again the post-perihelion time frame was easiest to characterize, and we confirmed the decrease in the apparent period first noted by Schleicher et al. (1990) and showed that it continued to decrease into June. This is overall consistent with the expected synodic effect for an object with pro-grade "rotation" – in Halley's case of non-principal axis motion it is the direction of the precession that is the same as the orbital motion, hence pro-grade – as the synodic period or solar "day" asymptotically approached the apparent sidereal period as Halley receded from the Sun. Quite unexpectedly, we identified a disjointed, stair-step behavior within the general decrease. Because these "steps" coincide with the timing of the change from a 3-peaked appearance to 2-peaked, we conclude that this is yet another consequence of the non-commensurability of the component periods. Though the periods are much less certain before perihelion, a similar downward step also takes place at the 3-to-2 transition. We also find evidence for generally higher values for the period prior to perihelion, but without an obvious explanation.

The beat frequency between the component periods directly implies an offset of about 0.45 day from a 2:1 ratio and, doubling the assumed 3.7 day precession period, yields a rotation period of ~6.95 or ~7.85 day. Can we identify which value is correct? Our intuition suggests that the apparent period that we measure should be intermediate between the rotation period and twice the precession period (with the source having moved somewhat more than 360° in one dimension and somewhat less than 360° in the other dimension to be closest to its original position). Since the synodic effect significantly perturbs this (primarily changing the effective precession period with respect to the Sun for an object that has a small obliquity), we examine the apparent period near the end of the apparition when the synodic period approaches its asymptotic limit, the apparent sidereal period. Since we measure a value somewhat less than 7.3 day, the shorter of the two options for the rotation component should be correct if the precession period is 3.7 day. However, for instance if the actual precession period is only 3.6 day and double yields 7.2 day, then an apparent asymptotic value just under 7.3 day implies that the rotation period is likely to be larger than the 2:1 ratio, thereby yielding ~7.65 day once the offset of 0.45 day is applied. Given the uncertainties, we conclude that neither possibility can as yet be ruled out. Finally, we suspect that the downward direction of the stair-steps is also associated with whether the ratio of component periods is greater or less than 2:1, but a resolution to this hypothesis also awaits detailed modeling.

### *4.2 Number and Locations of Sources*

Another obvious question is how many source regions are needed to explain the variations in the lightcurve and where are they located on the surface of the nucleus? Schleicher et al. (1990) assumed that multiple lightcurve features required multiple sources, while Belton et al. (1991) argued that five sources, including two equally strong sources, were needed to explain the various data sets. Crifo et al. (2002) and Szegö et al. (2002) have even claimed that there are no isolated source regions but that the entire surface is uniformly active and surface topography focuses outflowing material into denser features that are perceived as jets; however, the large lightcurve amplitudes readily eliminate this possibility. In their model, the maximum amplitude variation is determined by the change in cross-section of the nucleus, a factor of two for Halley. Yet the $C_3$ amplitude is more than a factor of four, and since the amplitudes increase as the aperture sizes decrease, we estimate the variation at the nucleus must be more than a factor of six. We, therefore, conclude that Halley's activity *must* be driven by isolated source regions, as most researchers have assumed.

Perhaps the most surprising finding is that a *single* source might explain all of the features during the entire post-perihelion time frame. Using the preferred complex scenario of both Belton et al. (1991) and Samarasinha & A'Hearn (1991), early simulations by N. Samarasinha (personal comm.) revealed that a single source could not only move in and out of sunlight multiple times in 7+ days, but that the illumination function would evolve back and forth in a manner consistent with the 3-to-2-to-3 peak pattern we observe. Our own more recent preliminary tests with our jet model confirm this finding and further reveal that more than one morphological feature can be visible at the same time due to projection effects coupled with the complex motion. The possibility that a single source region on Halley's nucleus might explain all or nearly all of the lightcurve features and their evolution following perihelion is remarkable and contrary to all expectations. If true, what about the multiple jets seen arising from different locations on the nucleus as imaged by Vega 2 and Giotto? It is possible that only one of these regions dominates the bulk gas and dust production and the "rotational" variability and other regions are only minor players, or have sufficiently similar effective longitudes that they turn on and off together.

The observed variability eliminates either end of the nucleus as the location of the primary source region because the amount of sunlight received by such a source at the rotational pole will be independent of the roll about the long axis, and so would yield a simple, equally spaced, double-peaked lightcurve. Since we do not see such a simple signature, we can rule out a source at either pole. We can also rule out the equator since, when averaged over the beat frequency, an equatorial source would receive an equal amount of sunlight independent of the Sun's orientation. Instead, we see a strong seasonal effect with differing bulk production rates and differing lightcurve amplitudes. Note that this is difficult to obtain with a complex rotator because no location on the surface ever goes into "winter," and made even more difficult if there are multiple source regions.

If we are correct that a single source dominates the activity following perihelion, then we conclude that this source region must systematically receive less illumination early in the apparition. The effective obliquity of the nucleus (i.e. the tilt of the total angular momentum axis with respect to the perpendicular of the orbital plane) is only 17° in the nominal solution of Belton et al. (1991) and Samarasinha & A'Hearn (1991), and so the average maximum altitude of the Sun as seen from the source region will only change by at most 34°. Moreover, the average sub-solar latitude varies by much less than this, only 12° from two months prior to two months following perihelion. Thus to cause a >50% increase in production rates, we conclude that the Sun is usually at low altitudes where a 12° change can have a proportionally greater change in the intensity.

### *4.3 Comparisons to Non-photometric Data Sets*

In addition to assisting our understanding of the detailed evolution of the lightcurve, our creation of a synthetic lightcurve has made it easier to compare the lightcurve to some other types of data. The most obvious is that of the dust and gas features observed in Halley's coma, and the first comparison of the April lightcurve with CN jets was performed by Hoban et al. (1988). Because Samarasinha et al. (in preparation) is currently reanalyzing the entire IHW archive of imaging to extract measurements of the jet morphology, we defer a more detailed inter-comparison. A related comparison is possible using the sequence of CN shells observed by Schulz and Schlosser (1988). By measuring the projected distance of the outward moving shells as a function of time, they extrapolated to an onset time at the nucleus for each feature, and we have overlaid these times onto our lightcurve in Figure 5. As is immediately evident, and as we would expect, each shell is associated with a minimum in the lightcurve when activity has resumed, or with a shoulder where outgassing has turned on but that more, older material continues to leave the aperture than is being created. Schulz (1992) subsequently showed that the CN shells, seen primarily at large distances from the nucleus, are associated with and a natural consequence of the spiral shaped CN jets discovered by A'Hearn et al. (1986); we note that it is much easier to derive start times from the shells than from the jets, and values from the jets themselves have not yet been determined. As an aside, an initial investigation into the correlation of the lightcurve with morphological features has recently been completed for the only other comet definitively known to be in a complex rotation – 103P/Hartley 2 (Knight et al. 2015).

[FIGURE 5 HERE]

In contrast to the strong correlation of the CN shells with minima in the lightcurve, we did not expect to find a similar correlation for disconnection events in Halley's ion tail since these have long been regarded as due to interactions with the solar wind, specifically the time of crossings past the comet by the heliocentric current sheet (cf. Brandt et al. 1999 and references therein). The location of the sheet, specifically its ecliptic latitude in the longitudinal direction and distance of the comet, and the times of its crossing of the comet, is usually not measured but must be modeled based on a variety of other data. As Brandt et al. show, some disconnection event (DE) start times directly coincide with apparent crossing times, others are close and adjustments in the model might cause a crossing, and a few are more problematic. Similar to CN shells, start times are based on observed distances and times for disconnections observed within the plasma tail, by either assuming or measuring both the acceleration and initial velocity for each event (see Brosius et al. 1987). Using the calculated start times for the DEs presented in Table 1 of Brandt et al., we also overlaid these values onto the lightcurve (Figure 5). The correlation of DEs with our lightcurve is clear, with ten of the twelve DEs observed after perihelion (numbers 8 through 19) coinciding with strong minima in our lightcurve. Of the remaining two, one (DE 16 at +51.2 day) occurs at a shoulder while the other and final event (DE 19 at +63.2 day) is about a day prior to the lightcurve minimum.

We note that the DEs are usually detected and followed down the tail during an interval of about 0.5-2.0 days after their extrapolated onset times. Each identified DE was measured on between one and 18 images, yielding estimated uncertainties of between less than an hour to one-half of a day. As is evident, several strong minima do not have an associated DE, but in most cases no images were acquired during the appropriate interval due to gaps in wide-field coverage, although at a couple of minima images exist and no DE was seen. Is it even possible that a decrease in the outgassing rate could cause a break in the plasma tail? The answer appears to be yes. Brandt et al. (1999) indicate that the plasma tail itself disappeared relatively abruptly during the first week of May and did not return. Average water production rates at this time were only about a factor of two below the production rates at lightcurve minima in early March, and we've already noted that an extrapolation of amplitudes vs aperture size suggest that gas production rates *at the nucleus* likely varied by more than six-fold. Since the production rates at strong minima are comparable to the rates when the entire plasma tail disappeared, we conclude that such a drop could cause the observed breaks. In fact, Wurm & Mammano (1972) argued that an unknown mechanism intrinsic to the comet might be the cause of these and other plasma tail events. In conclusion, while we don't argue that all DEs are caused by this mechanism, the observed correlation certainly implies that intrinsic cometary activity must be an additional method along with current sheet crossings.

### *4.4 Expected Lightcurve Behavior at Large Heliocentric Distances*

Several attempts were made to determine Halley's rotation period when it was still far from the Sun in 1984 and 1985, with limited success. Festou et al.'s (1987) and Belton's (1990) reanalyses of these data found evidence for 7+ day periodicity but were not conclusive regarding the nature of the lightcurve, while Mueller & Belton (1993) found evidence for a double-peaked 3.7 day period in 1984 January. In fact, we expect considerable differences depending on if brightness measurements at larger heliocentric distances are dominated by the coma or by the nucleus. For the coma driven case, we anticipate much smaller amplitudes because parent (and daughter) lifetimes are much longer than near perihelion. We would expect the 3-to-2-to-3 peak oscillation to continue and the apparent period to vary with a saw-tooth shape about the apparent sidereal period of slightly less than 7.3 day, with sharp drops followed by a slow rise every 6 to 8 weeks. In contrast, if the active region(s) have not yet turned on (or turned off when the comet is outbound), then we expect to see a quadruple-peaked, sinusoidal lightcurve with exactly twice the precession period, and having a nearly 2x amplitude due to Halley's cross-sectional variation. In other words, it would look like a typical double-peaked asteroidal lightcurve with the precessional period, with only the lightcurve shape perhaps slightly modified by the rotation about the long axis due to Halley not being axial symmetric.

## 5. CONCLUSIONS

In spite of the somewhat inhomogeneous nature of the component datasets, particularly a wide range of aperture sizes used during data acquisition, the inclusion and analysis of the entire IHW photometry archive has resulted in a number of discoveries. Not only does the lightcurve shape constantly evolve, it does so in a regular manner that must be associated with the beat frequency of its component periods and can be used to tightly constrain the rotational component once the precessional component, near 3.7 day, is precisely determined. The measured apparent period not only exhibits the expected pro-grade synodic variation as the length of a solar "day" varies as Halley approached and receded from the Sun, but also shows a stair-step or saw-tooth variation strongly correlated with the evolution of the lightcurve shape. The strong seasonal asymmetry in production rates, coupled with the small effective obliquity of the total angular momentum vector, implies the need for relatively low illumination angles of the source region(s), while preliminary modeling indicates that a *single* source might produce the entire post-perihelion lightcurve variations and associated evolution. As expected, initiation times for gas jets are directly matched to local minima in the lightcurve, but quite unexpectedly, we also find a strong correlation of plasma tail disconnection events with the same lightcurve minima, suggesting that a sufficient drop in outgassing can cause breaks in the ion tail. Two quite different predictions regarding the characteristics of the lightcurve at larger heliocentric distances are made, depending on whether the coma or the nucleus dominates the measurements.

Following the companion analysis of the inner coma morphology from IHW archive images by Samarasinha et al. (in prep.), our next combined project will be to perform jet modeling similar to what we have carried out for other comets such as Hyakutake (1992 B2) (Schleicher & Woodney 2003). Because Halley's complex motion requires many more free parameters than when modeling a body in simple rotation, exploring the full parameter space would normally be nearly hopeless. It is only with the many additional observational constraints identified above and those imposed by groundbased coma morphology, along with the snapshots obtained during the spacecraft flybys, that we can hope to unravel Halley's many riddles.

## ACKNOWLEDGEMENTS

We particularly thank M. A'Hearn for making available the IHW photometry database in a useful form shortly after its creation and for assistance in resolving problems within the database. The data were subsequently incorporated into the Small Bodies Node of the Planetary Data System. We thank M. A'Hearn and N. Samarasinha for numerous useful discussions regarding non-principal axis rotation and other aspects of the data and modeling, and R. Millis for starting us down this path and inspiration along the way. This work would not have been possible without the observations and contributions from numerous observers around the world who made the Halley campaign a successful endeavor. Many of the lightcurve analyses were performed using the DataDesk software from Data Descriptions, Inc. This research was supported by NASA's Planetary Atmospheres Program (grant NNX11AD85G), students L. Alciatore Stinnett and R. Williams were supported through NSF REU grant AST-9200137 at Northern Arizona University, and students S. Sackey and B. Smith-Konter were supported by the NASA Space Grant program at Northern Arizona University.

## REFERENCES

A'Hearn, M. F. 1991. In "The Comet Halley Archive Summary Volume", ed. Z. Sekanina (Pasadena, CA; JPL), 193.

A'Hearn, M. F., & Carsenty, U. 1992. COMET HALLEY PHOTOMETRIC FLUXES V1.0, IHW-C-PPFLX-3-RDR-HALLEY-V1.0, NASA Planetary Data System.

A'Hearn, M. F., Hoban, S., Birch, P. V., et al. 1986. Nature 324 649.

A'Hearn, M. F., Millis, R. L., Schleicher, D. G., Osip, D. J., & Birch, P. V. 1995. Icarus 118, 223.

A'Hearn, M. F., Schleicher, D. G., Feldman, P. D., Millis, R. L., & Thompson, D. T. 1984. AJ 89, 579.

A'Hearn, M. F., & Vanysek, V. 2006. IHW COMET HALLEY PHOTOMETRIC FLUXES V2.0, IHW-C-PPFLX-3-RDR-HALLEY-V2.0, NASA Planetary Data System.

Belton, M. J. S. 1990, Icarus 86, 30.

Belton, M. J. S., Julian, W. H., Anderson, A. J., & Mueller, B. E. A. 1991, Icarus 93, 183.

Brandt, J. C., Caputo, F. M., Hoeksema, J. T., et al. 1999. Icarus 137, 69.

Brosius, J. W., et al. 1987. A&A 187, 267.

Crifo, J.-F., Rodionov, A. V., Szegö, K., & Fulle, M. 1992. Earth, Moon, & Planets 90, 227.

Cruikshank, D. P., Hartmann, W. K., & Tholen, D. J. 1985. Nature 315, 122.

Feldman, P. D., Festou, M. C., A'Hearn, M. F., et al. 1987. A&A 187, 325.

Festou, M. C., Drossart, P., Lecacheux, J., et al. 1987. A&A 187, 575.

Hoban, S., Samarasinha, N. H., A'Hearn, M. F., & Klinglesmith, D. A. 1988. A&A 195, 331.

Keller, H. U., Arpigny, C., Barbieri, C., et al. 1986. Nature 321, 320.

Knight, M. M., Mueller, B. E. A., Samarasinha, N. H., & Schleicher, D. G. 2015. AJ, in press.

McFadden, L. A., A'Hearn, M. F., Feldman, P. D., et al. 1987. A&A 187, 333.

Millis, R. L., A'Hearn, M. F., & Campins, H. 1985. BAAS 17, 688.

Millis, R. L., & Schleicher, D. G. 1986. Nature 324, 646.

Mueller, B. E. A., & Belton, M. J. S. 1993. In "Proceedings of the Workshop on the Activity of Distant Comets", ed. W. F. Huebner, H. U. Keller, D. Jewitt, J. Klinger, & R. West (San Antonio, TX; Southwest Research Inst.), 21.

Neckel, T., & Münch, G. 1987. A&A 187, 581.

Osborn, W., A'Hearn, M. F., Carsenty, U., et al. 1990. Icarus 88, 228.

Sagdeev, R. A., Szabó, F., Avanesov, G. A., et al. 1986. Nature 321, 262.

Samarasinha, N. H., & A'Hearn, M. F. 1991, Icarus 93, 194.

Samarasinha, N. H., A'Hearn, M. F., Hoban, S., & Klinglesmith, D. A. 1986. ESA SP-250. Exploration of Halley's Comet 1, 487.

Schleicher, D. G., & Bus, S. J. 1991. AJ 101, 706.

Schleicher, D. G., Millis, R. L., & Birch, P. V. 1998, Icarus 132, 397.

Schleicher, D. G., Millis, R. L., Thompson, D. T., et al. 1990. AJ 100, 896.

Schleicher, D. G., & Woodney, L. M. 2003. Icarus 162, 190.

Schulz, R. 1992. Icarus 96, 198.

Schulz, R., & Schlosser, W. 1989. A&A 214, 375.

Sekanina, Z., & Larson, S. M. 1984, AJ 89, 1408.

Sekanina, Z., & Larson, S. M. 1986, AJ 92, 462.

Szegö, K., Crifo, J.-F., Rodionov, A. V., & Fulle, M. 1992. Earth, Moon, & Planets 90, 435.

Wurm, K., & Mammano, A. 1972. Astrophys. & Space Sc. 18, 273.

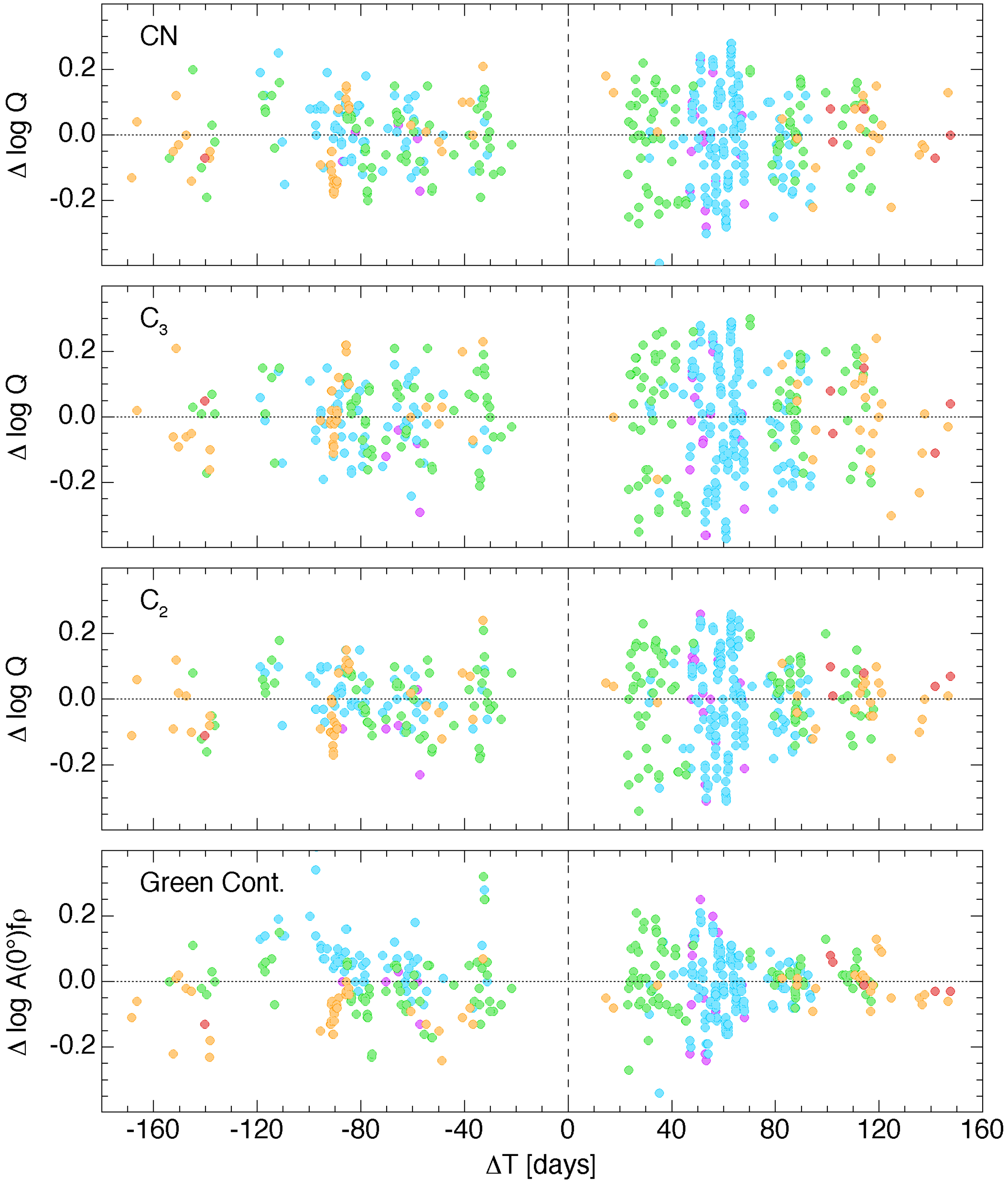

**Figure 1.** Differential production rates, following the removal of secular trends, as a function of time from perihelion. The most useful gas species — CN, $C_3$, and $C_2$ — are presented as well as the best continuum bandpass — green at 4845Å — in these four panels. Each plotted point is the average of all data obtained in ~2 hr bins for each individual telescope. Points are color-coded by the mean projected aperture radius, ρ: <1x$10^4$ km (purple), 1-2x$10^4$ km (blue), 2-4x$10^4$ km (green), 4-8x$10^4$ km (orange), and >8x$10^4$ km (red). Readily evident are trends associated with the envelope of the lightcurve variations with time, aperture size, and species.

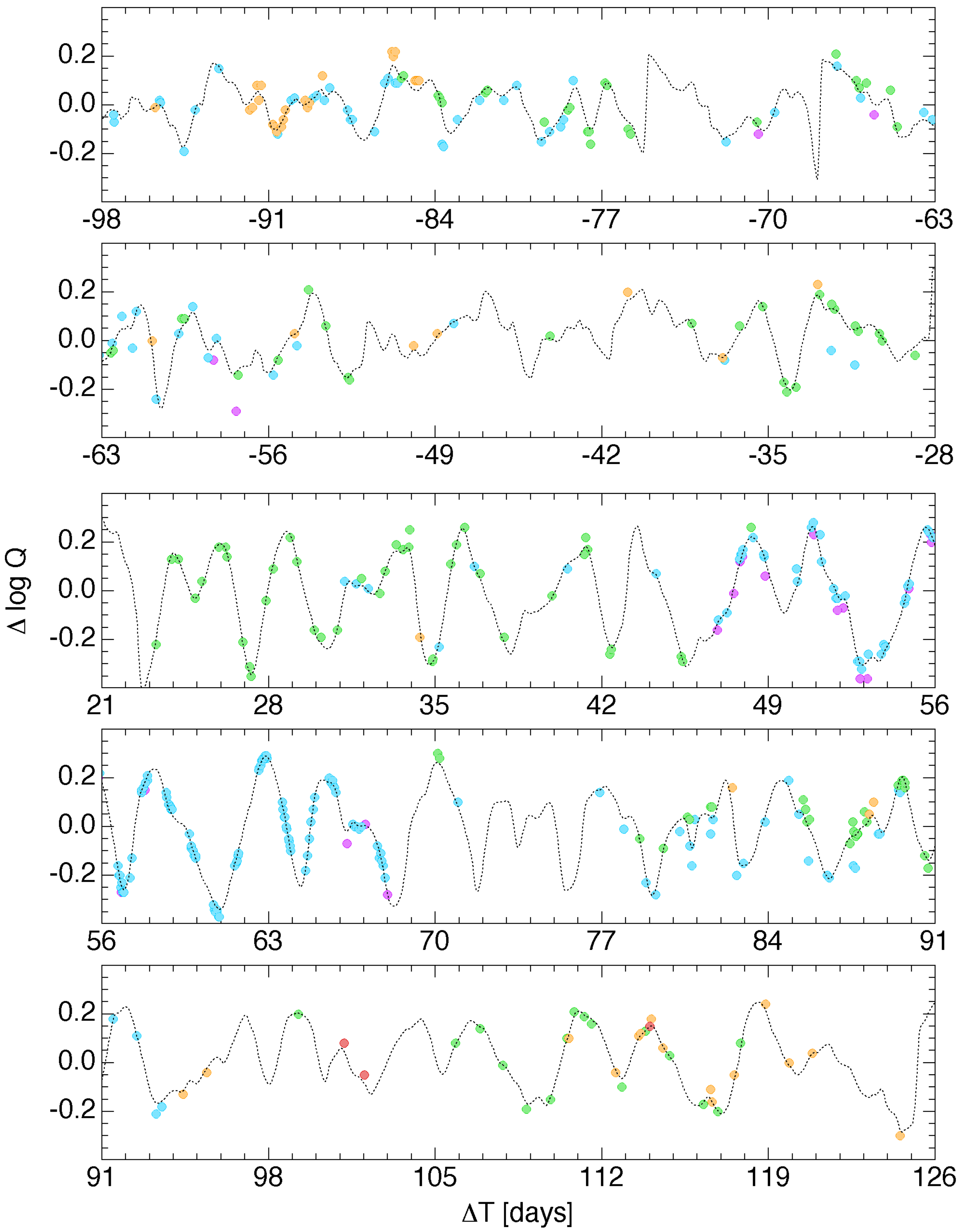

**Figure 2.** Differential production rates of $C_3$ as a function of time from perihelion. Each panel gives a 35-day interval, allowing the entire, constantly evolving lightcurve to be examined. Points are shown with the same color-coded projected aperture size ranges as Figure 1. Data are overlaid with a synthetic curve that was based on the basic 7+ day periodicity and the characteristics of the slow evolution seen from cycle to cycle (see Section 3.3).

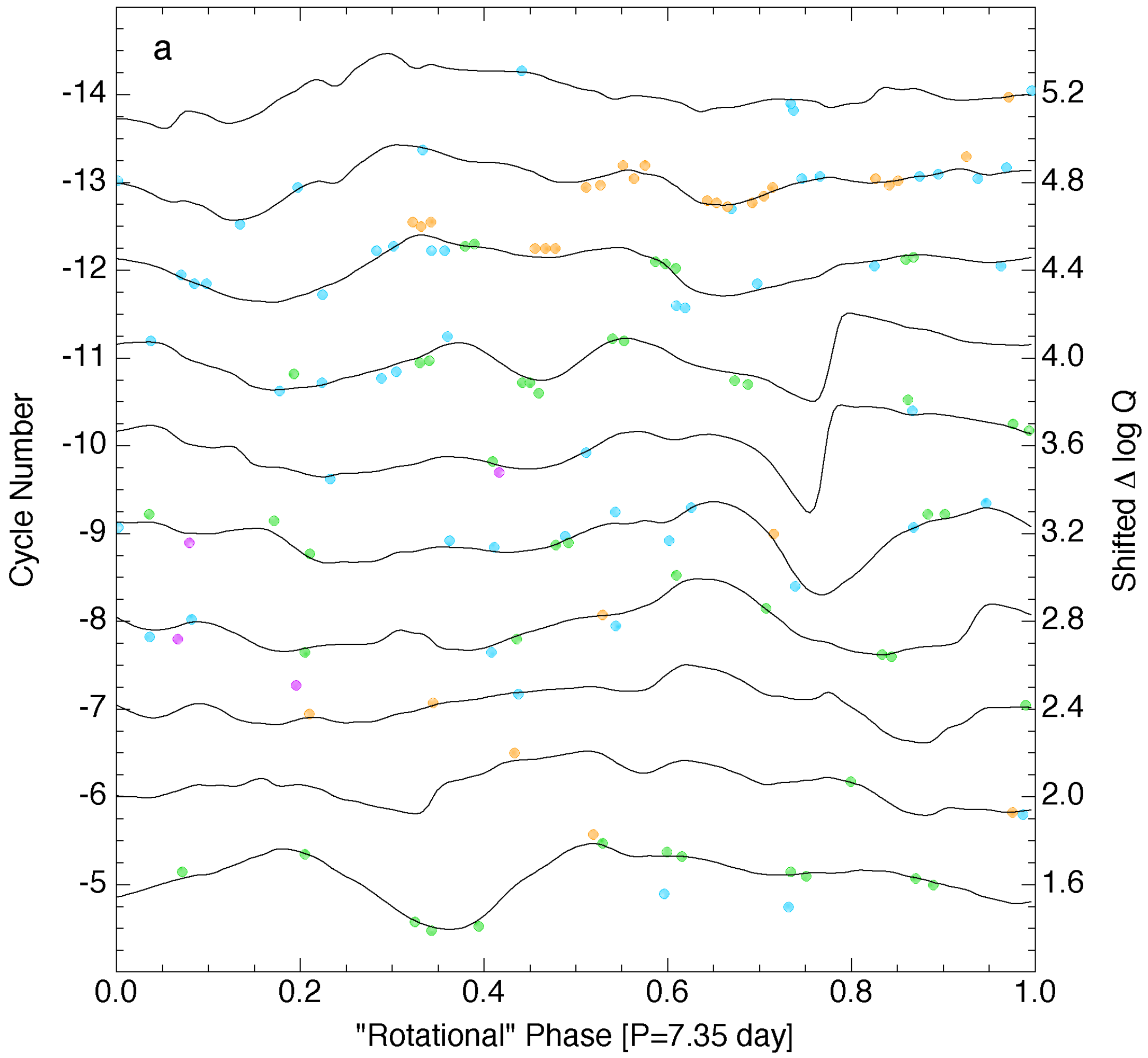

**Figure 3.** Differential production rates of $C_3$ as a function of rotational phase. Pre-perihelion $C_3$ data (panel "a") and post-perihelion $C_3$ data (panel "b") are shown with the same color-coded projected aperture size ranges as Figure 1 and phased by the average apparent period of 7.35 day, with each successive "rotational" cycle shifted lower in Δ log *Q* by 0.4. Phasing is normalized to the time of perihelion (1986 Feb 9.459) and progresses from cycle –14 to cycle –5 before perihelion, corresponding to an interval of just over 73 days, and after perihelion from cycle 4 to cycle 17, corresponding to an interval of just over 102 days. The data are overlaid by our synthetic lightcurve that represents our best estimate of the evolving lightcurve shape based solely on the measured data points, the period, and the systematic slow evolution at every rotational phase (see Section 3.3). As is evident from the synthetic curve, some features, such as the post-perihelion minimum near phase 0.75 and the maximum near phase 0.90, persist for many cycles while other features substantially change shape in only a few weeks time.

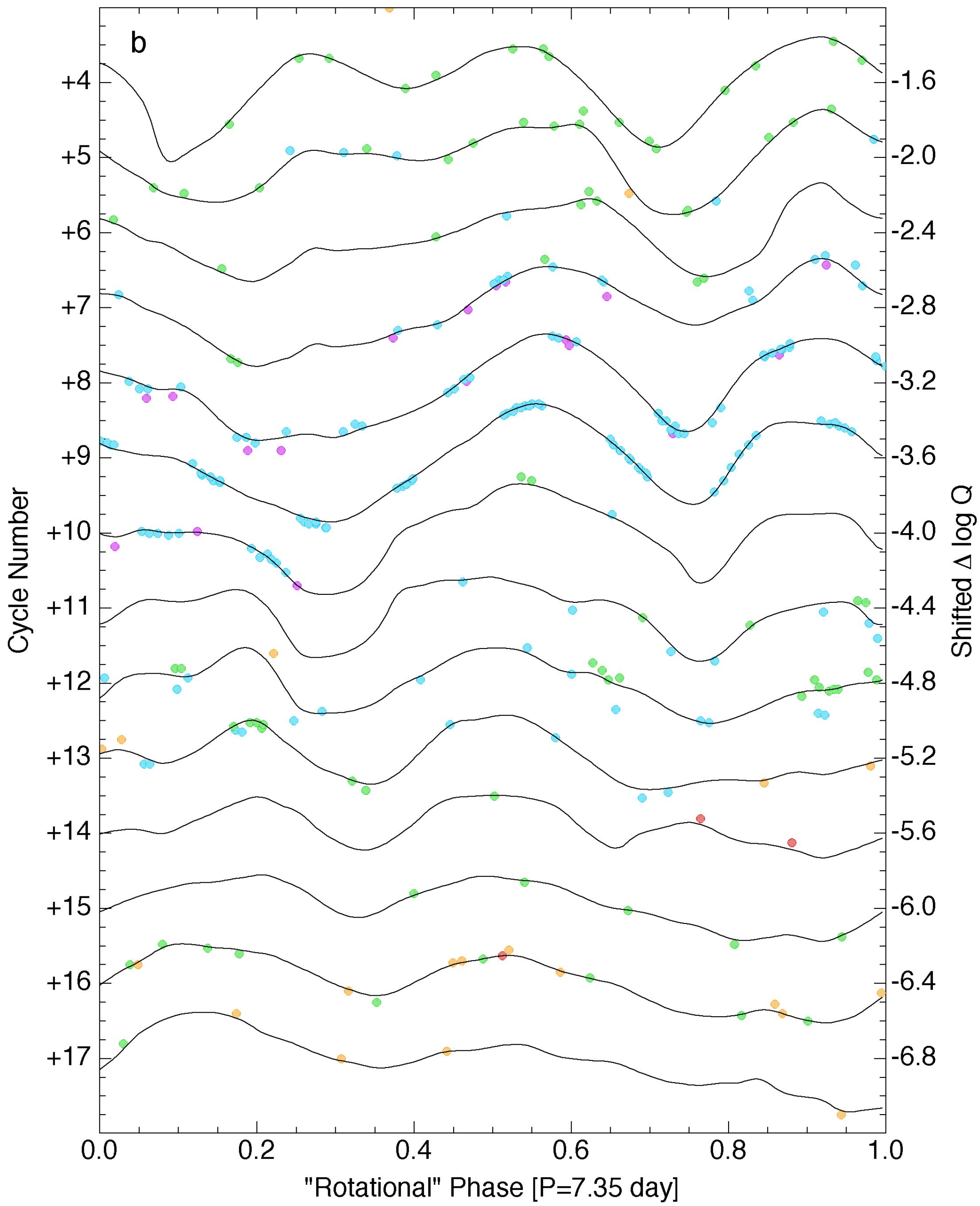

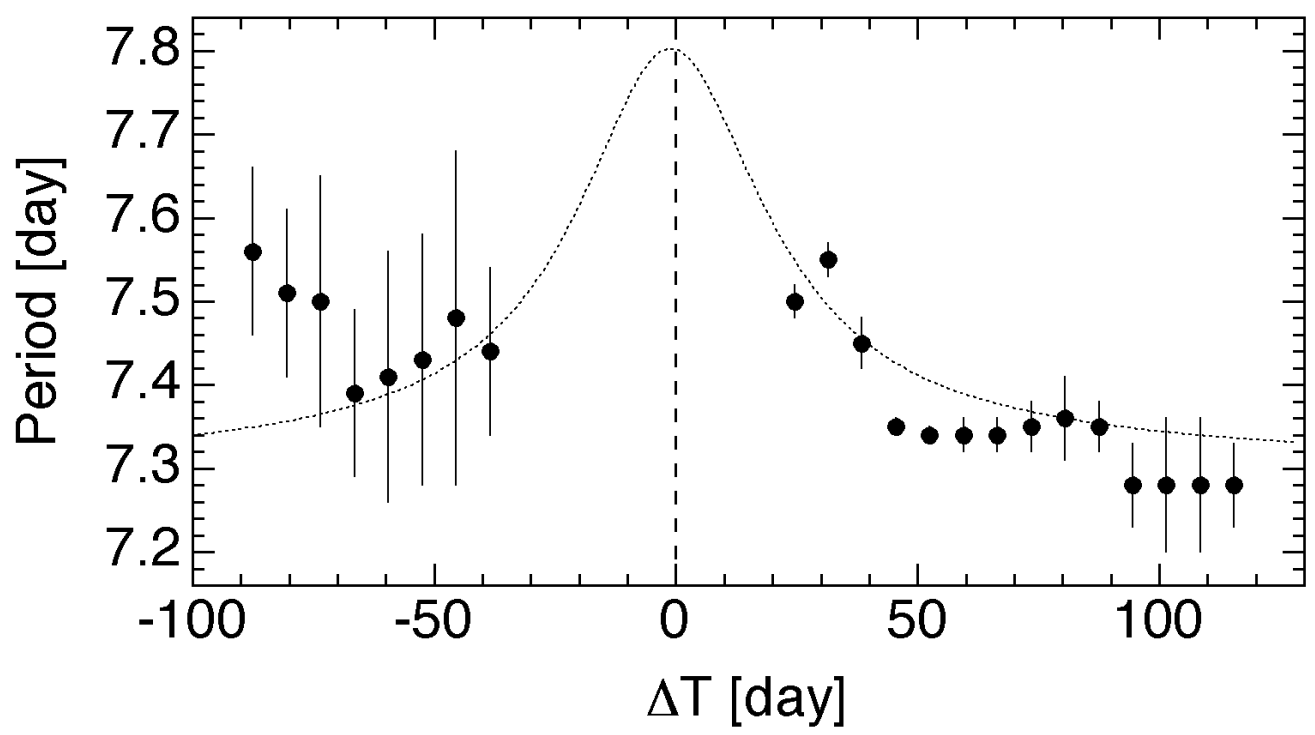

**Figure 4.** Derived apparent periods from $C_3$ as a function of time from perihelion. Each derived period is based on a 3-week interval centered at the plotted time from perihelion and measured every 7 days. Overlaid for reference is a synodic, i.e. solar, period curve based on using the preferred total angular momentum vector solution derived both by Belton et al. (1991) and Samarasinha & A'Hearn (1991) and treating it as the rotational axis of a simple rotator. Note the asymmetry about perihelion and the quasi-stair step type of behavior; we think both of these attributes are caused by the timing of when a source region on Halley's nucleus receives sunlight due the underlying component precession and rotation periods being slightly non-commensurate.

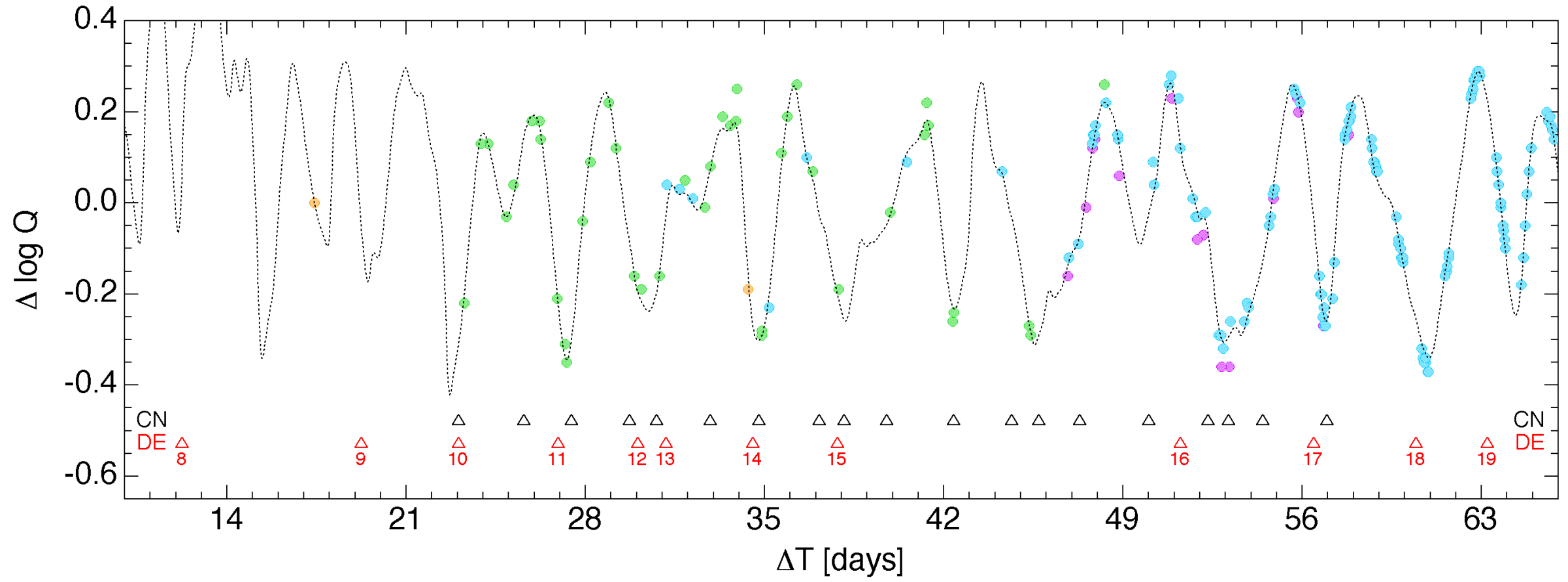

**Figure 5.** The time of events as compared to the photometric lightcurve of $C_3$. Post-perihelion $C_3$ data and the corresponding synthetic curve are plotted in the same manner as in Figure 2. Also plotted are the extrapolated start times for CN shells (black triangles) derived by Schulz & Schlosser (1989); as is evident, each of the onset times correspond to either a minimum in the lightcurve or to a shoulder when activity has increased (a shoulder occurs when there is more, older material moving out of the aperture than is being newly released from the nucleus). We also plot the extrapolated start times for disconnection events (red, numbered triangles) observed in Halley's ion tail as derived by Brandt et al. (1999); these times also exhibit a strong correlation with minima in the lightcurve, implying that major decreases in outgassing can play a significant role in the creation of disconnection events.

**Table 1**
Site Information

| Observatory | Tel ID[a] | Telescope Used | Number of Nights | Number of binned sets[b] | Submitter[c] |
|---|---|---|---|---|---|
| Lowell Observatory, USA | 1 | 1.8 m | 3 | 3 | D. Schleicher |
| Lowell Observatory, USA | 2 | 1.1 m F/16 | 13 | 13 | D. Schleicher |
| Lowell Observatory, USA | 3 | 1.1 m F/8 | 2 | 3 | D. Schleicher |
| Lowell Observatory, USA | 5 | 0.6 m | 1 | 1 | D. Schleicher |
| Cerro Tololo Inter-American Observatory, Chile | 6 | 0.6 m | 48 | 110 | D. Schleicher |
| Perth Observatory, Australia | 9 | 0.6 m | 34 | 54 | P. Birch |
| Mount John University Observatory, New Zealand | 10 | 0.6 m | 13 | 14 | J. Manfroid |
| European Southern Observatory, Chile | 15 | 0.5 m | 10 | 11 | J. Manfroid |
| Cerro Tololo Inter-American Observatory, Chile | 16 | 1.0 m | 13 | 35 | D. Schleicher |
| Mauna Kea Observatory, USA | 17 | 0.6 m | 59 | 63 | D. J. Tholen |
| Mt Lemmon Observatory, USA | 18 | 1.5 m | 5 | 5 | W. Wisniewski |
| Lick Observatory, USA | 19 | 0.6 m | 3 | 3 | R. P. S. Stone |
| Bosque Alegre Astrophysical Station, Argentina | 20 | 1.5 m | 10 | 14 | J. J. Claria |
| La Palma Observatory, Canary Islands, Spain | 21 | 1.0 m | 9 | 11 | A. Fitzsimmons |
| Catania Astrophysical Observatory, Italy | 22 | 0.9 m | 22 | 36 | G. Strazzulla |
| Jena Observatory, Germany | 23 | 0.9 m | 3 | 5 | W. Pfau |
| South African Astronomical Observatory, South Africa | 25 | 0.5 m | 5 | 5 | P. J. Andrews |
| Wise Observatory, Israel | 26 | 1.0 m | 2 | 3 | J. Manfroid |
| Mt. Sanglok Observatory, Tajikistan (formerly USSR) | 27 | 1.0 m | 50 | 57 | N. N. Kiselev |
| Tarija Observatory, Bolivia | 28 | 0.6 m | 10 | 13 | N. N. Kiselev |
| Tarija Observatory, Bolivia | 29 | 0.6 m | 1 | 2 | N. N. Kiselev |

[a]The number assigned to each telescope for our analyses
[b]The number of binned data sets, which average ~2 hr of data, from each telescope
[c]Submitter of data to the IHW archive

**Table 2**
Production Rate $r_H$–Dependence for Comet 1P/Halley

| | | Second Order Fit[a] | | | |
|---|---|---|---|---|---|
| Species | Interval | A(0) | A(1) | A(2) | $\chi^2$ |
| OH | pre | 29.31 | −1.71 | −1.61 | 0.006 |
| | post | 29.40 | −1.60 | −0.87 | 0.015 |
| CN | pre | 26.86 | −3.19 | 0.08 | 0.010 |
| | post | 27.01 | −2.38 | 1.09 | 0.018 |
| $C_3$ | pre | 25.97 | −2.39 | −1.57 | 0.011 |
| | post | 26.15 | −2.10 | 0.55 | 0.029 |
| $C_2$ | pre | 27.08 | −2.84 | −2.70 | 0.008 |
| | post | 27.19 | −2.26 | −0.43 | 0.019 |
| Dust | pre | 4.32 | −3.86 | 2.11 | 0.012 |
| | post | 4.65 | −2.67 | 1.54 | 0.009 |

[a] Second order fits of log $Q$(X) (or log $A(0°)f\rho$) with log $r_H$.

**Table 3**
Residual Photometric Production Rates for Comet 1P/Halley

| UT Date | | ΔT | Tel | # | Mean | Residual Production Rates[e] | | | |
|---|---|---|---|---|---|---|---|---|---|
| | | (day)[a] | ID[b] | Obs[c] | log ρ[d] | CN | $C_3$ | $C_2$ | Green |
| 1985 | | | | | | | | | |
| Aug | 24.95 | –168.510 | 27 | 1 | 4.71 | –0.13 | — | –0.11 | –0.11 |
| Aug | 26.96 | –166.499 | 27 | 1 | 4.70 | 0.04 | 0.02 | 0.06 | –0.06 |
| Sep | 8.40 | –154.059 | 2 | 1 | 4.58 | –0.07 | — | — | 0.00 |
| Sep | 9.98 | –152.476 | 27 | 1 | 4.64 | –0.05 | –0.06 | –0.09 | –0.22 |
| Sep | 10.96 | –151.496 | 27 | 2 | 4.64 | 0.12 | 0.21 | 0.12 | 0.01 |
| Sep | 11.97 | –150.490 | 27 | 2 | 4.63 | –0.03 | –0.09 | 0.02 | 0.02 |
| Sep | 14.96 | –147.498 | 27 | 4 | 4.62 | 0.00 | –0.06 | 0.01 | –0.02 |
| Sep | 16.95 | –145.506 | 27 | 4 | 4.61 | –0.14 | –0.05 | –0.10 | –0.03 |
| Sep | 17.43 | –145.029 | 1 | 1 | 4.39 | 0.20 | 0.03 | 0.08 | 0.11 |
| Sep | 20.92 | –141.538 | 27 | 3 | 4.58 | –0.10 | 0.01 | –0.12 | –0.02 |
| Sep | 22.06 | –140.400 | 26 | 4 | 4.91 | –0.07 | 0.05 | –0.11 | –0.13 |
| Sep | 22.93 | –139.533 | 27 | 4 | 4.57 | –0.19 | –0.17 | –0.16 | –0.04 |
| Sep | 23.99 | –138.463 | 26 | 4 | 4.90 | –0.07 | –0.16 | –0.08 | –0.23 |
| Sep | 24.07 | –138.394 | 26 | 3 | 4.90 | –0.05 | –0.10 | –0.05 | –0.18 |
| Sep | 24.94 | –137.520 | 27 | 3 | 4.56 | 0.03 | 0.07 | –0.05 | 0.03 |
| Sep | 25.97 | –136.487 | 27 | 5 | 4.55 | –0.02 | 0.01 | –0.08 | 0.00 |
| Oct | 13.44 | –119.016 | 1 | 2 | 4.30 | 0.19 | 0.06 | 0.10 | 0.13 |
| Oct | 14.50 | –117.960 | 17 | 2 | 4.50 | 0.12 | 0.15 | 0.06 | 0.05 |
| Oct | 15.34 | –117.122 | 1 | 1 | 4.20 | 0.12 | –0.01 | 0.08 | 0.14 |
| Oct | 15.46 | –117.003 | 17 | 1 | 4.49 | 0.08 | 0.01 | 0.04 | 0.03 |
| Oct | 15.58 | –116.881 | 17 | 1 | 4.49 | 0.07 | 0.01 | 0.02 | 0.06 |
| Oct | 18.01 | –114.454 | 27 | 2 | 4.38 | 0.12 | 0.12 | 0.12 | 0.07 |
| Oct | 19.02 | –113.443 | 27 | 1 | 4.37 | –0.04 | –0.14 | 0.05 | –0.07 |
| Oct | 20.46 | –111.995 | 18 | 1 | 4.13 | 0.25 | 0.14 | 0.10 | 0.19 |
| Oct | 20.97 | –111.492 | 27 | 2 | 4.35 | 0.16 | 0.15 | 0.18 | 0.15 |
| Oct | 22.03 | –110.433 | 23 | 3 | 4.15 | –0.02 | –0.14 | –0.08 | 0.14 |
| Oct | 22.99 | –109.460 | 23 | 2 | 4.14 | –0.15 | — | — | 0.14 |
| Nov | 1.80 | –99.660 | 27 | 3 | 4.23 | 0.08 | 0.11 | 0.07 | 0.20 |
| Nov | 3.95 | –97.507 | 27 | 3 | 4.20 | 0.03 | –0.04 | 0.05 | 0.14 |
| Nov | 3.97 | –97.488 | 23 | 4 | 4.01 | 0.08 | –0.07 | –0.03 | 0.34 |
| Nov | 4.05 | –97.407 | 23 | 4 | 4.01 | 0.00 | — | –0.03 | 0.41 |
| Nov | 5.69 | –95.765 | 9 | 2 | 4.73 | –0.09 | –0.01 | –0.09 | –0.15 |
| Nov | 5.88 | –95.582 | 27 | 4 | 4.18 | 0.08 | 0.02 | 0.09 | 0.11 |
| Nov | 5.92 | –95.540 | 27 | 1 | 4.18 | 0.09 | 0.01 | 0.10 | 0.10 |
| Nov | 6.90 | –94.564 | 27 | 4 | 4.17 | –0.10 | –0.19 | –0.03 | 0.07 |
| Nov | 7.36 | –94.100 | 2 | 2 | 4.17 | 0.08 | –0.02 | –0.07 | 0.07 |
| Nov | 8.36 | –93.104 | 2 | 1 | 4.23 | 0.19 | 0.15 | 0.10 | 0.10 |
| Nov | 9.66 | –91.795 | 9 | 4 | 4.68 | –0.09 | –0.02 | –0.10 | –0.13 |
| Nov | 9.78 | –91.681 | 9 | 4 | 4.68 | –0.09 | –0.01 | –0.10 | –0.13 |
| Nov | 9.96 | –91.503 | 22 | 3 | 4.62 | –0.05 | 0.08 | 0.00 | –0.07 |
| Nov | 10.05 | –91.414 | 22 | 3 | 4.62 | –0.11 | 0.02 | –0.05 | –0.12 |
| Nov | 10.14 | –91.324 | 22 | 2 | 4.62 | –0.05 | 0.08 | –0.01 | –0.06 |
| Nov | 10.63 | –90.828 | 9 | 2 | 4.66 | –0.13 | –0.08 | –0.14 | –0.16 |
| Nov | 10.71 | –90.753 | 9 | 4 | 4.66 | –0.15 | –0.09 | –0.16 | –0.16 |
| Nov | 10.80 | –90.663 | 9 | 3 | 4.66 | –0.17 | –0.11 | –0.17 | –0.16 |
| Nov | 10.83 | –90.633 | 27 | 2 | 4.12 | –0.01 | –0.12 | –0.01 | 0.04 |
| Nov | 10.99 | –90.468 | 22 | 4 | 4.61 | –0.18 | –0.09 | –0.10 | –0.09 |
| Nov | 11.08 | –90.375 | 22 | 3 | 4.61 | –0.14 | –0.06 | –0.07 | –0.10 |
| Nov | 11.16 | –90.302 | 22 | 3 | 4.61 | –0.17 | –0.02 | –0.11 | –0.12 |
| Nov | 11.39 | –90.070 | 17 | 1 | 4.20 | 0.02 | 0.02 | –0.02 | 0.05 |
| Nov | 11.53 | –89.927 | 17 | 1 | 4.20 | 0.03 | 0.03 | 0.01 | 0.07 |
| Nov | 11.98 | –89.482 | 22 | 6 | 4.60 | –0.15 | 0.02 | –0.09 | –0.07 |
| Nov | 12.09 | –89.370 | 22 | 3 | 4.60 | –0.14 | –0.01 | –0.08 | –0.08 |
| Nov | 12.16 | –89.300 | 22 | 3 | 4.60 | –0.14 | 0.01 | –0.09 | –0.08 |
| Nov | 12.33 | –89.127 | 17 | 1 | 4.19 | 0.02 | 0.03 | — | 0.06 |

| | | | | | | | | | | |
|---|---|---|---|---|---|---|---|---|---|---|
| Nov | 12.48 | –88.979 | 17 | 1 | 4.19 | 0.07 | 0.04 | — | 0.10 |
| Nov | 12.71 | –88.753 | 9 | 4 | 4.64 | 0.08 | 0.12 | 0.08 | –0.05 |
| Nov | 12.80 | –88.661 | 27 | 3 | 4.10 | 0.07 | 0.02 | 0.03 | 0.06 |
| Nov | 13.02 | –88.435 | 27 | 1 | 4.10 | 0.12 | 0.07 | 0.09 | 0.09 |
| Nov | 13.78 | –87.683 | 27 | 3 | 4.09 | 0.01 | –0.02 | 0.04 | 0.08 |
| Nov | 13.88 | –87.578 | 27 | 3 | 4.09 | –0.02 | –0.06 | 0.03 | 0.06 |
| Nov | 13.98 | –87.483 | 27 | 3 | 4.09 | –0.03 | –0.06 | 0.02 | 0.07 |
| Nov | 14.25 | –87.212 | 19 | 1 | 3.88 | –0.08 | — | –0.09 | 0.00 |
| Nov | 14.90 | –86.555 | 27 | 2 | 4.08 | –0.08 | –0.11 | –0.04 | 0.03 |
| Nov | 15.34 | –86.121 | 17 | 1 | 4.16 | 0.09 | 0.09 | 0.08 | 0.09 |
| Nov | 15.47 | –85.985 | 17 | 1 | 4.16 | 0.10 | 0.11 | 0.15 | 0.16 |
| Nov | 15.63 | –85.832 | 9 | 2 | 4.61 | 0.11 | 0.22 | 0.10 | –0.04 |
| Nov | 15.70 | –85.764 | 9 | 4 | 4.61 | 0.14 | 0.20 | 0.12 | –0.03 |
| Nov | 15.77 | –85.689 | 9 | 3 | 4.61 | 0.15 | 0.22 | 0.15 | 0.00 |
| Nov | 15.78 | –85.682 | 27 | 3 | 4.07 | 0.06 | 0.09 | 0.10 | 0.16 |
| Nov | 15.88 | –85.576 | 27 | 2 | 4.07 | 0.09 | 0.09 | 0.12 | 0.16 |
| Nov | 16.05 | –85.412 | 22 | 11 | 4.35 | 0.06 | 0.11 | 0.08 | 0.06 |
| Nov | 16.12 | –85.341 | 22 | 6 | 4.37 | 0.05 | 0.12 | 0.08 | 0.05 |
| Nov | 16.61 | –84.853 | 9 | 3 | 4.60 | 0.09 | 0.10 | 0.09 | –0.02 |
| Nov | 16.69 | –84.772 | 9 | 4 | 4.60 | 0.09 | 0.10 | 0.09 | –0.02 |
| Nov | 16.77 | –84.693 | 9 | 4 | 4.60 | 0.08 | 0.10 | 0.11 | –0.04 |
| Nov | 17.57 | –83.892 | 9 | 2 | 4.59 | 0.06 | 0.04 | 0.08 | –0.04 |
| Nov | 17.65 | –83.812 | 9 | 4 | 4.59 | 0.05 | 0.03 | 0.08 | –0.06 |
| Nov | 17.73 | –83.729 | 9 | 3 | 4.59 | 0.05 | 0.01 | 0.07 | –0.05 |
| Nov | 17.74 | –83.723 | 27 | 2 | 4.05 | –0.07 | –0.16 | –0.01 | 0.06 |
| Nov | 17.81 | –83.653 | 27 | 2 | 4.05 | –0.08 | –0.17 | –0.02 | 0.05 |
| Nov | 18.38 | –83.077 | 17 | 2 | 4.13 | –0.01 | –0.06 | — | 0.06 |
| Nov | 19.26 | –82.200 | 18 | 1 | 3.81 | 0.01 | — | — | 0.00 |
| Nov | 19.31 | –82.141 | 2 | 1 | 4.11 | 0.07 | 0.02 | –0.01 | 0.00 |
| Nov | 19.57 | –81.890 | 9 | 2 | 4.58 | 0.02 | 0.05 | 0.03 | –0.10 |
| Nov | 19.63 | –81.826 | 9 | 3 | 4.58 | 0.02 | 0.06 | 0.03 | –0.10 |
| Nov | 20.33 | –81.128 | 2 | 2 | 4.19 | 0.09 | 0.02 | –0.02 | –0.03 |
| Nov | 20.88 | –80.577 | 27 | 1 | 4.02 | 0.09 | 0.08 | 0.15 | 0.08 |
| Nov | 21.91 | –79.546 | 27 | 2 | 4.02 | –0.12 | –0.15 | 0.07 | 0.05 |
| Nov | 22.03 | –79.431 | 22 | 6 | 4.32 | –0.10 | –0.07 | –0.02 | –0.05 |
| Nov | 22.25 | –79.210 | 2 | 2 | 4.17 | –0.02 | –0.11 | –0.04 | 0.00 |
| Nov | 22.73 | –78.734 | 27 | 2 | 4.01 | –0.05 | –0.09 | –0.03 | 0.00 |
| Nov | 22.84 | –78.615 | 27 | 2 | 4.01 | –0.02 | –0.06 | –0.01 | 0.02 |
| Nov | 23.03 | –78.426 | 22 | 6 | 4.31 | –0.05 | –0.02 | –0.04 | –0.05 |
| Nov | 23.11 | –78.352 | 22 | 6 | 4.31 | –0.04 | –0.01 | –0.03 | –0.06 |
| Nov | 23.25 | –78.206 | 2 | 2 | 4.09 | 0.18 | 0.10 | 0.03 | 0.06 |
| Nov | 23.85 | –77.608 | 22 | 6 | 4.31 | –0.18 | –0.11 | –0.07 | –0.11 |
| Nov | 23.92 | –77.543 | 22 | 6 | 4.31 | –0.17 | –0.11 | –0.07 | –0.10 |
| Nov | 23.99 | –77.474 | 22 | 6 | 4.31 | –0.20 | –0.16 | –0.07 | –0.11 |
| Nov | 24.58 | –76.883 | 9 | 3 | 4.55 | 0.04 | 0.09 | 0.05 | –0.04 |
| Nov | 24.67 | –76.792 | 9 | 4 | 4.55 | 0.04 | 0.08 | 0.05 | –0.02 |
| Nov | 25.55 | –75.907 | 9 | 3 | 4.54 | –0.10 | –0.10 | –0.08 | –0.23 |
| Nov | 25.66 | –75.801 | 9 | 4 | 4.54 | –0.13 | –0.12 | –0.11 | –0.22 |
| Nov | 29.67 | –71.791 | 27 | 2 | 4.00 | –0.11 | –0.15 | –0.04 | 0.08 |
| Nov | 30.96 | –70.496 | 22 | 6 | 4.31 | –0.06 | –0.07 | –0.06 | –0.03 |
| Dec | 1.02 | –70.439 | 22 | 2 | 3.98 | –0.06 | –0.12 | –0.09 | 0.00 |
| Dec | 1.71 | –69.747 | 27 | 2 | 4.02 | 0.02 | –0.03 | –0.02 | 0.04 |
| Dec | 4.29 | –67.172 | 3 | 3 | 4.42 | 0.15 | 0.21 | 0.10 | 0.05 |
| Dec | 4.32 | –67.136 | 17 | 2 | 4.11 | 0.12 | 0.16 | 0.06 | 0.12 |
| Dec | 5.13 | –66.330 | 3 | 6 | 4.37 | 0.07 | 0.10 | 0.04 | 0.05 |
| Dec | 5.25 | –66.205 | 3 | 4 | 4.37 | 0.06 | 0.07 | 0.05 | 0.04 |
| Dec | 5.32 | –66.135 | 17 | 2 | 4.12 | 0.01 | 0.03 | 0.00 | 0.07 |
| Dec | 5.57 | –65.892 | 9 | 4 | 4.54 | 0.04 | 0.09 | 0.05 | –0.01 |
| Dec | 5.89 | –65.568 | 23 | 4 | 3.85 | 0.03 | –0.04 | –0.08 | 0.03 |
| Dec | 6.57 | –64.892 | 9 | 4 | 4.59 | 0.05 | 0.06 | 0.04 | –0.03 |
| Dec | 6.86 | –64.604 | 22 | 6 | 4.35 | –0.07 | –0.09 | –0.05 | –0.07 |
| Dec | 7.97 | –63.487 | 21 | 1 | 4.20 | –0.07 | –0.03 | –0.10 | –0.01 |

| | | | | | | | | | | |
|---|---|---|---|---|---|---|---|---|---|---|
| Dec | 8.33 | –63.129 | 17 | 2 | 4.10 | –0.04 | –0.06 | –0.09 | 0.00 |
| Dec | 8.82 | –62.636 | 22 | 12 | 4.37 | –0.08 | –0.05 | –0.10 | –0.09 |
| Dec | 8.89 | –62.565 | 21 | 2 | 4.13 | –0.02 | –0.01 | –0.09 | –0.03 |
| Dec | 8.92 | –62.537 | 22 | 6 | 4.37 | –0.06 | –0.04 | –0.08 | –0.07 |
| Dec | 9.30 | –62.163 | 17 | 1 | 4.15 | 0.08 | 0.10 | 0.01 | 0.02 |
| Dec | 9.73 | –61.730 | 27 | 2 | 4.07 | 0.11 | –0.03 | 0.09 | 0.08 |
| Dec | 9.90 | –61.556 | 21 | 4 | 4.24 | 0.08 | 0.12 | — | 0.04 |
| Dec | 10.56 | –60.894 | 9 | 3 | 4.62 | 0.03 | 0.00 | 0.02 | –0.09 |
| Dec | 10.74 | –60.722 | 27 | 2 | 4.08 | –0.13 | –0.24 | –0.09 | –0.06 |
| Dec | 11.68 | –59.777 | 27 | 2 | 4.09 | 0.05 | 0.03 | 0.03 | 0.06 |
| Dec | 11.80 | –59.663 | 22 | 6 | 4.40 | 0.01 | 0.09 | 0.00 | –0.01 |
| Dec | 11.93 | –59.527 | 22 | 7 | 4.43 | 0.04 | 0.09 | 0.03 | 0.02 |
| Dec | 12.26 | –59.198 | 17 | 1 | 4.18 | 0.12 | 0.14 | 0.07 | 0.18 |
| Dec | 12.92 | –58.537 | 22 | 2 | 4.07 | –0.11 | –0.07 | –0.04 | 0.02 |
| Dec | 13.15 | –58.313 | 19 | 1 | 3.90 | –0.01 | –0.08 | 0.03 | 0.00 |
| Dec | 13.26 | –58.200 | 17 | 1 | 4.19 | 0.01 | 0.01 | –0.01 | 0.07 |
| Dec | 14.09 | –57.365 | 18 | 1 | 3.89 | –0.17 | –0.29 | –0.23 | –0.13 |
| Dec | 14.16 | –57.296 | 2 | 3 | 4.34 | –0.08 | –0.14 | –0.12 | –0.06 |
| Dec | 15.66 | –55.803 | 27 | 2 | 4.13 | –0.08 | –0.14 | –0.06 | –0.07 |
| Dec | 15.86 | –55.602 | 22 | 9 | 4.35 | –0.11 | –0.08 | –0.11 | –0.16 |
| Dec | 16.55 | –54.912 | 9 | 2 | 4.68 | 0.01 | 0.03 | –0.02 | –0.13 |
| Dec | 16.65 | –54.808 | 27 | 2 | 4.14 | 0.01 | –0.02 | –0.01 | –0.03 |
| Dec | 17.14 | –54.322 | 5 | 2 | 4.38 | 0.15 | 0.21 | 0.12 | 0.05 |
| Dec | 17.85 | –53.607 | 22 | 10 | 4.35 | 0.03 | 0.06 | 0.08 | 0.05 |
| Dec | 18.78 | –52.678 | 22 | 6 | 4.47 | –0.17 | –0.15 | –0.16 | –0.17 |
| Dec | 18.86 | –52.603 | 22 | 6 | 4.47 | –0.16 | –0.16 | –0.15 | –0.17 |
| Dec | 21.55 | –49.911 | 9 | 2 | 4.73 | –0.02 | –0.02 | –0.04 | –0.15 |
| Dec | 22.54 | –48.920 | 9 | 2 | 4.74 | –0.05 | 0.03 | –0.12 | –0.24 |
| Dec | 23.22 | –48.239 | 17 | 1 | 4.29 | 0.00 | 0.07 | –0.02 | 0.01 |
| Dec | 27.28 | –44.181 | 17 | 1 | 4.32 | –0.06 | 0.02 | –0.08 | –0.01 |
| Dec | 30.54 | –40.920 | 9 | 2 | 4.80 | 0.10 | 0.20 | 0.08 | –0.11 |
| 1986 | | | | | | | | | |
| Jan | 2.23 | –38.228 | 17 | 1 | 4.37 | 0.00 | 0.07 | 0.03 | 0.05 |
| Jan | 2.53 | –37.929 | 9 | 1 | 4.83 | 0.10 | — | 0.07 | –0.08 |
| Jan | 3.53 | –36.933 | 9 | 1 | 4.83 | 0.00 | –0.07 | –0.06 | –0.13 |
| Jan | 3.61 | –36.853 | 27 | 2 | 4.29 | –0.11 | –0.08 | –0.09 | –0.08 |
| Jan | 4.23 | –36.225 | 17 | 1 | 4.38 | –0.03 | 0.06 | –0.02 | –0.06 |
| Jan | 5.21 | –35.244 | 17 | 1 | 4.39 | 0.08 | 0.14 | 0.09 | 0.06 |
| Jan | 6.23 | –34.233 | 17 | 2 | 4.40 | –0.11 | –0.21 | –0.18 | –0.07 |
| Jan | 6.10 | –34.364 | 2 | 1 | 4.38 | –0.09 | –0.17 | –0.15 | –0.02 |
| Jan | 6.60 | –33.856 | 27 | 2 | 4.31 | –0.19 | –0.19 | –0.17 | –0.13 |
| Jan | 7.12 | –33.339 | 2 | 1 | 4.39 | 0.09 | — | 0.05 | 0.04 |
| Jan | 7.13 | –33.324 | 19 | 1 | 4.11 | 0.11 | — | 0.03 | 0.11 |
| Jan | 7.52 | –32.937 | 9 | 2 | 4.86 | 0.21 | 0.23 | 0.24 | 0.07 |
| Jan | 7.59 | –32.864 | 27 | 1 | 4.32 | 0.11 | 0.19 | 0.21 | 0.32 |
| Jan | 8.09 | –32.370 | 18 | 1 | 4.09 | 0.04 | –0.04 | 0.09 | 0.28 |
| Jan | 8.11 | –32.349 | 2 | 4 | 4.36 | 0.14 | 0.15 | 0.13 | 0.25 |
| Jan | 8.23 | –32.230 | 17 | 1 | 4.41 | 0.13 | 0.13 | 0.08 | 0.25 |
| Jan | 9.08 | –31.375 | 18 | 1 | 4.10 | –0.06 | –0.10 | –0.06 | 0.00 |
| Jan | 9.10 | –31.358 | 2 | 3 | 4.46 | 0.06 | 0.06 | 0.02 | 0.06 |
| Jan | 9.23 | –31.234 | 17 | 1 | 4.41 | –0.01 | 0.04 | –0.01 | 0.02 |
| Jan | 10.10 | –30.358 | 2 | 3 | 4.47 | 0.01 | 0.03 | –0.03 | –0.06 |
| Jan | 10.24 | –30.220 | 17 | 1 | 4.42 | –0.04 | 0.00 | –0.03 | –0.09 |
| Jan | 11.60 | –28.855 | 27 | 1 | 4.34 | –0.12 | –0.06 | –0.02 | –0.09 |
| Jan | 14.58 | –25.880 | 27 | 2 | 4.35 | –0.11 | –0.06 | –0.06 | –0.07 |
| Jan | 18.59 | –21.869 | 27 | 2 | 4.37 | –0.03 | –0.03 | 0.08 | –0.02 |
| Feb | 23.87 | 14.414 | 9 | 3 | 4.81 | 0.18 | — | 0.05 | –0.05 |
| Feb | 26.86 | 17.405 | 9 | 3 | 4.79 | 0.13 | 0.00 | 0.04 | –0.08 |
| Mar | 4.37 | 22.914 | 6 | 3 | 4.55 | –0.12 | — | –0.17 | –0.07 |
| Mar | 4.72 | 23.262 | 10 | 1 | 4.53 | –0.25 | –0.22 | –0.26 | –0.27 |
| Mar | 4.85 | 23.387 | 9 | 2 | 4.52 | 0.04 | — | 0.00 | –0.02 |
| Mar | 5.37 | 23.912 | 6 | 3 | 4.52 | 0.09 | 0.13 | 0.10 | 0.00 |
| Mar | 5.65 | 24.194 | 17 | 1 | 4.36 | 0.09 | 0.13 | 0.14 | 0.10 |
| Mar | 6.37 | 24.908 | 6 | 3 | 4.51 | 0.03 | –0.03 | 0.03 | 0.01 |
| Mar | 6.65 | 25.191 | 17 | 1 | 4.35 | –0.01 | 0.04 | 0.04 | 0.03 |
| Mar | 7.37 | 25.909 | 6 | 3 | 4.56 | 0.17 | 0.18 | 0.16 | 0.11 |
| Mar | 7.65 | 26.195 | 17 | 1 | 4.34 | 0.13 | 0.18 | 0.17 | 0.21 |
| Mar | 7.71 | 26.248 | 10 | 3 | 4.51 | 0.09 | 0.14 | 0.07 | 0.10 |
| Mar | 8.36 | 26.905 | 6 | 4 | 4.51 | 0.00 | –0.21 | –0.05 | 0.09 |
| Mar | 8.65 | 27.188 | 17 | 1 | 4.34 | –0.22 | –0.31 | –0.24 | –0.01 |
| Mar | 8.71 | 27.253 | 10 | 2 | 4.49 | –0.27 | –0.35 | –0.34 | –0.07 |
| Mar | 9.35 | 27.895 | 6 | 4 | 4.54 | –0.07 | –0.04 | –0.08 | –0.06 |
| Mar | 9.64 | 28.181 | 17 | 1 | 4.32 | 0.00 | 0.09 | 0.05 | 0.01 |
| Mar | 10.36 | 28.906 | 6 | 5 | 4.49 | 0.22 | 0.22 | 0.23 | 0.16 |
| Mar | 10.64 | 29.177 | 17 | 1 | 4.32 | 0.11 | 0.12 | 0.16 | 0.18 |
| Mar | 11.36 | 29.902 | 6 | 5 | 4.43 | –0.02 | –0.16 | –0.05 | 0.01 |
| Mar | 11.65 | 30.187 | 17 | 1 | 4.30 | –0.16 | –0.19 | –0.15 | –0.08 |
| Mar | 12.35 | 30.893 | 6 | 6 | 4.42 | –0.18 | –0.16 | –0.21 | –0.18 |
| Mar | 12.64 | 31.180 | 17 | 1 | 4.30 | –0.07 | 0.04 | –0.01 | –0.10 |
| Mar | 13.14 | 31.678 | 25 | 4 | 4.27 | 0.01 | 0.03 | 0.06 | –0.01 |
| Mar | 13.36 | 31.896 | 6 | 7 | 4.44 | 0.05 | 0.05 | 0.08 | –0.03 |
| Mar | 13.64 | 32.179 | 17 | 1 | 4.28 | –0.01 | 0.01 | 0.02 | –0.03 |
| Mar | 14.12 | 32.659 | 25 | 5 | 4.34 | –0.01 | –0.01 | 0.02 | –0.06 |
| Mar | 14.35 | 32.888 | 6 | 6 | 4.47 | 0.07 | 0.08 | 0.08 | –0.02 |
| Mar | 14.82 | 33.361 | 9 | 6 | 4.60 | 0.16 | 0.19 | 0.16 | 0.05 |
| Mar | 15.11 | 33.647 | 25 | 5 | 4.33 | 0.13 | 0.17 | 0.17 | 0.19 |
| Mar | 15.34 | 33.883 | 6 | 6 | 4.46 | 0.17 | 0.18 | 0.18 | 0.16 |
| Mar | 15.38 | 33.922 | 20 | 3 | 4.52 | 0.13 | 0.25 | 0.15 | 0.15 |
| Mar | 15.80 | 34.345 | 9 | 3 | 4.63 | 0.01 | –0.19 | –0.01 | –0.01 |
| Mar | 16.34 | 34.884 | 6 | 7 | 4.44 | –0.20 | –0.29 | –0.24 | –0.07 |
| Mar | 16.36 | 34.898 | 20 | 5 | 4.50 | –0.24 | –0.28 | –0.23 | –0.05 |
| Mar | 16.62 | 35.161 | 17 | 1 | 4.25 | –0.39 | –0.23 | –0.27 | –0.34 |
| Mar | 17.11 | 35.652 | 25 | 5 | 4.30 | –0.01 | 0.11 | 0.05 | 0.02 |
| Mar | 17.34 | 35.883 | 6 | 6 | 4.36 | 0.11 | 0.19 | 0.13 | 0.10 |
| Mar | 17.70 | 36.241 | 10 | 4 | 4.40 | 0.13 | 0.26 | 0.14 | 0.12 |
| Mar | 18.09 | 36.634 | 25 | 5 | 4.29 | 0.10 | 0.10 | 0.14 | 0.12 |
| Mar | 18.34 | 36.877 | 6 | 6 | 4.43 | 0.12 | 0.07 | 0.11 | 0.09 |
| Mar | 19.35 | 37.890 | 20 | 6 | 4.36 | –0.21 | –0.19 | –0.12 | –0.07 |
| Mar | 21.35 | 39.893 | 20 | 5 | 4.38 | –0.08 | –0.02 | 0.03 | –0.09 |
| Mar | 22.01 | 40.552 | 27 | 1 | 4.09 | –0.03 | 0.09 | 0.11 | 0.02 |
| Mar | 22.70 | 41.246 | 10 | 11 | 4.33 | 0.04 | 0.15 | 0.04 | 0.08 |
| Mar | 22.78 | 41.322 | 9 | 4 | 4.49 | 0.14 | 0.22 | 0.15 | 0.10 |
| Mar | 22.86 | 41.397 | 9 | 5 | 4.40 | 0.08 | 0.17 | 0.15 | 0.08 |
| Mar | 23.79 | 42.335 | 9 | 3 | 4.43 | –0.21 | –0.26 | –0.22 | –0.08 |
| Mar | 23.85 | 42.396 | 9 | 4 | 4.48 | –0.20 | –0.24 | –0.22 | –0.10 |
| Mar | 25.73 | 44.274 | 10 | 1 | 4.28 | –0.01 | 0.07 | –0.15 | 0.03 |
| Mar | 26.78 | 45.323 | 9 | 7 | 4.34 | –0.20 | –0.27 | –0.20 | –0.12 |
| Mar | 26.85 | 45.391 | 9 | 4 | 4.35 | –0.21 | –0.29 | –0.23 | –0.12 |
| Mar | 28.30 | 46.842 | 15 | 2 | 3.89 | –0.17 | –0.16 | –0.15 | –0.22 |
| Mar | 28.34 | 46.886 | 16 | 6 | 4.24 | –0.16 | –0.12 | –0.15 | –0.18 |
| Mar | 28.71 | 47.255 | 10 | 11 | 4.24 | –0.19 | –0.09 | –0.21 | –0.20 |
| Mar | 28.99 | 47.540 | 27 | 2 | 3.98 | –0.05 | –0.01 | 0.00 | –0.07 |
| Mar | 29.24 | 47.783 | 6 | 2 | 4.09 | 0.07 | 0.13 | 0.08 | 0.03 |
| Mar | 29.27 | 47.806 | 15 | 2 | 3.87 | 0.08 | 0.12 | 0.11 | 0.08 |
| Mar | 29.29 | 47.834 | 16 | 6 | 4.22 | 0.05 | 0.15 | 0.07 | –0.01 |
| Mar | 29.33 | 47.874 | 6 | 8 | 4.20 | 0.09 | 0.15 | 0.09 | 0.05 |
| Mar | 29.35 | 47.894 | 15 | 2 | 3.87 | 0.10 | 0.14 | 0.13 | 0.11 |
| Mar | 29.37 | 47.910 | 16 | 7 | 4.22 | 0.07 | 0.17 | 0.10 | 0.02 |
| Mar | 29.72 | 48.261 | 9 | 5 | 4.41 | 0.15 | 0.26 | 0.17 | 0.11 |
| Mar | 29.79 | 48.335 | 9 | 3 | 4.27 | 0.13 | 0.22 | 0.17 | 0.14 |
| Mar | 30.25 | 48.790 | 6 | 5 | 4.25 | 0.12 | 0.15 | 0.15 | 0.13 |
| Mar | 30.27 | 48.811 | 16 | 3 | 4.20 | 0.10 | 0.14 | 0.16 | 0.15 |
| Mar | 30.30 | 48.840 | 15 | 1 | 3.91 | 0.06 | 0.06 | 0.12 | 0.14 |
| Mar | 30.61 | 49.147 | 17 | 1 | 4.04 | — | — | 0.00 | 0.02 |
| Mar | 31.63 | 50.170 | 17 | 1 | 4.03 | –0.01 | 0.09 | –0.01 | –0.02 |
| Mar | 31.66 | 50.204 | 10 | 15 | 4.19 | –0.04 | 0.04 | –0.09 | –0.05 |
| Apr | 1.25 | 50.787 | 6 | 6 | 4.17 | 0.22 | 0.26 | 0.22 | 0.21 |
| Apr | 1.34 | 50.883 | 6 | 7 | 4.15 | 0.24 | 0.28 | 0.24 | 0.21 |
| Apr | 1.35 | 50.891 | 15 | 2 | 3.82 | 0.23 | 0.23 | 0.26 | 0.25 |
| Apr | 1.62 | 51.164 | 17 | 1 | 4.01 | 0.20 | 0.23 | 0.23 | 0.19 |
| Apr | 1.69 | 51.229 | 10 | 23 | 4.17 | 0.16 | 0.12 | 0.10 | 0.18 |
| Apr | 2.18 | 51.721 | 6 | 2 | 4.15 | 0.08 | 0.01 | 0.09 | 0.06 |
| Apr | 2.28 | 51.819 | 6 | 8 | 4.12 | 0.04 | –0.03 | 0.04 | 0.01 |
| Apr | 2.36 | 51.897 | 6 | 4 | 4.20 | 0.04 | –0.03 | 0.03 | 0.03 |
| Apr | 2.34 | 51.886 | 15 | 2 | 3.80 | 0.00 | –0.08 | 0.02 | –0.03 |
| Apr | 2.59 | 52.130 | 17 | 2 | 3.99 | –0.02 | –0.07 | –0.04 | –0.05 |
| Apr | 2.67 | 52.209 | 10 | 24 | 4.16 | –0.09 | –0.02 | –0.06 | –0.03 |
| Apr | 3.19 | 52.732 | 6 | 5 | 4.16 | –0.16 | –0.29 | –0.20 | –0.12 |
| Apr | 3.28 | 52.819 | 6 | 3 | 4.27 | –0.16 | –0.29 | –0.21 | –0.10 |
| Apr | 3.29 | 52.833 | 15 | 2 | 3.79 | –0.23 | –0.36 | –0.26 | –0.22 |
| Apr | 3.36 | 52.901 | 6 | 6 | 4.19 | –0.19 | –0.32 | –0.24 | –0.14 |
| Apr | 3.60 | 53.146 | 17 | 1 | 3.98 | –0.28 | –0.36 | –0.31 | –0.24 |
| Apr | 3.65 | 53.194 | 10 | 36 | 4.14 | –0.30 | –0.26 | –0.30 | –0.19 |
| Apr | 4.18 | 53.725 | 6 | 6 | 4.17 | –0.21 | –0.26 | –0.27 | –0.19 |
| Apr | 4.29 | 53.836 | 6 | 4 | 4.22 | –0.20 | –0.22 | –0.26 | –0.19 |
| Apr | 4.36 | 53.900 | 6 | 4 | 4.03 | –0.20 | –0.23 | –0.26 | –0.22 |
| Apr | 5.16 | 54.701 | 6 | 5 | 4.16 | –0.06 | –0.05 | –0.11 | –0.11 |
| Apr | 5.22 | 54.764 | 6 | 7 | 4.11 | –0.04 | –0.03 | –0.09 | –0.11 |
| Apr | 5.31 | 54.851 | 6 | 5 | 4.20 | –0.01 | 0.02 | –0.06 | –0.07 |
| Apr | 5.34 | 54.877 | 15 | 2 | 3.77 | 0.00 | 0.01 | 0.00 | –0.06 |
| Apr | 5.37 | 54.910 | 6 | 6 | 4.10 | 0.01 | 0.03 | –0.03 | –0.06 |
| Apr | 6.14 | 55.679 | 6 | 4 | 4.15 | 0.23 | 0.25 | 0.22 | 0.16 |
| Apr | 6.19 | 55.735 | 6 | 6 | 4.10 | 0.22 | 0.24 | 0.22 | 0.17 |
| Apr | 6.26 | 55.806 | 6 | 4 | 4.00 | 0.22 | 0.23 | 0.22 | 0.17 |
| Apr | 6.30 | 55.837 | 15 | 2 | 3.75 | 0.19 | 0.20 | 0.22 | 0.20 |
| Apr | 6.36 | 55.904 | 6 | 11 | 4.15 | 0.21 | 0.22 | 0.21 | 0.15 |
| Apr | 7.13 | 56.669 | 6 | 7 | 4.11 | –0.04 | –0.16 | –0.05 | –0.01 |
| Apr | 7.16 | 56.703 | 16 | 5 | 4.09 | –0.21 | –0.20 | –0.17 | –0.04 |
| Apr | 7.21 | 56.746 | 6 | 6 | 4.13 | –0.07 | –0.20 | –0.08 | –0.03 |
| Apr | 7.25 | 56.787 | 16 | 5 | 4.09 | –0.15 | –0.25 | –0.12 | –0.05 |
| Apr | 7.27 | 56.808 | 15 | 2 | 3.74 | –0.14 | –0.27 | –0.13 | –0.09 |
| Apr | 7.29 | 56.829 | 6 | 6 | 4.13 | –0.09 | –0.23 | –0.11 | –0.04 |
| Apr | 7.32 | 56.861 | 16 | 6 | 4.09 | –0.18 | –0.27 | –0.15 | –0.08 |
| Apr | 7.37 | 56.912 | 6 | 8 | 4.16 | –0.13 | –0.27 | –0.15 | –0.07 |
| Apr | 7.63 | 57.175 | 10 | 33 | 4.09 | –0.23 | –0.21 | –0.24 | –0.12 |
| Apr | 7.71 | 57.255 | 10 | 6 | 4.09 | –0.17 | –0.13 | –0.21 | –0.11 |
| Apr | 8.12 | 57.665 | 16 | 3 | 4.09 | 0.09 | 0.14 | 0.08 | 0.04 |
| Apr | 8.12 | 57.657 | 6 | 6 | 4.15 | 0.10 | 0.15 | 0.06 | 0.05 |
| Apr | 8.20 | 57.737 | 6 | 7 | 4.09 | 0.11 | 0.16 | 0.08 | 0.08 |
| Apr | 8.26 | 57.803 | 15 | 2 | 3.74 | 0.10 | 0.15 | 0.12 | 0.15 |
| Apr | 8.27 | 57.816 | 6 | 6 | 4.13 | 0.13 | 0.18 | 0.11 | 0.10 |
| Apr | 8.29 | 57.830 | 16 | 6 | 4.09 | 0.13 | 0.18 | 0.14 | 0.09 |
| Apr | 8.36 | 57.899 | 16 | 4 | 4.09 | 0.14 | 0.19 | 0.15 | 0.11 |
| Apr | 8.36 | 57.904 | 6 | 8 | 4.14 | 0.14 | 0.21 | 0.11 | 0.12 |
| Apr | 9.16 | 58.702 | 6 | 7 | 4.16 | 0.16 | 0.14 | 0.16 | 0.09 |
| Apr | 9.17 | 58.714 | 16 | 6 | 4.07 | 0.16 | 0.12 | 0.17 | 0.10 |
| Apr | 9.25 | 58.794 | 6 | 6 | 4.08 | 0.13 | 0.09 | 0.12 | 0.06 |
| Apr | 9.26 | 58.806 | 16 | 8 | 4.07 | 0.13 | 0.09 | 0.15 | 0.08 |
| Apr | 9.33 | 58.869 | 6 | 7 | 4.03 | 0.11 | 0.08 | 0.10 | 0.05 |
| Apr | 9.35 | 58.896 | 16 | 8 | 4.07 | 0.12 | 0.07 | 0.13 | 0.06 |
| Apr | 9.39 | 58.930 | 6 | 3 | 4.14 | 0.11 | 0.07 | 0.11 | 0.05 |
| Apr | 10.12 | 59.666 | 6 | 3 | 4.20 | 0.02 | –0.03 | 0.00 | 0.02 |
| Apr | 10.21 | 59.746 | 6 | 7 | 4.07 | –0.02 | –0.08 | –0.06 | –0.02 |
| Apr | 10.21 | 59.753 | 16 | 8 | 4.07 | –0.02 | –0.09 | –0.03 | –0.01 |
| Apr | 10.29 | 59.835 | 6 | 10 | 4.12 | –0.04 | –0.10 | –0.06 | –0.02 |
| Apr | 10.32 | 59.860 | 16 | 8 | 4.07 | –0.05 | –0.12 | –0.05 | –0.02 |
| Apr | 10.38 | 59.921 | 6 | 4 | 4.13 | –0.05 | –0.12 | –0.07 | –0.03 |
| Apr | 10.38 | 59.924 | 16 | 3 | 4.07 | –0.06 | –0.13 | –0.06 | –0.04 |
| Apr | 11.13 | 60.674 | 6 | 3 | 4.14 | –0.21 | –0.32 | –0.25 | –0.11 |
| Apr | 11.17 | 60.714 | 16 | 5 | 4.07 | –0.26 | –0.34 | –0.25 | –0.12 |
| Apr | 11.21 | 60.753 | 6 | 6 | 4.04 | –0.24 | –0.35 | –0.29 | –0.15 |
| Apr | 11.28 | 60.818 | 16 | 10 | 4.07 | –0.27 | –0.35 | –0.28 | –0.14 |
| Apr | 11.28 | 60.819 | 6 | 5 | 4.13 | –0.23 | –0.34 | –0.28 | –0.13 |
| Apr | 11.37 | 60.911 | 16 | 8 | 4.07 | –0.28 | –0.37 | –0.30 | –0.15 |
| Apr | 11.38 | 60.918 | 6 | 8 | 4.10 | –0.26 | –0.37 | –0.31 | –0.16 |
| Apr | 12.04 | 61.576 | 6 | 6 | 4.14 | –0.15 | –0.16 | –0.22 | –0.13 |
| Apr | 12.09 | 61.628 | 16 | 11 | 4.07 | –0.16 | –0.15 | –0.19 | –0.15 |
| Apr | 12.13 | 61.666 | 6 | 5 | 4.04 | –0.14 | –0.14 | –0.21 | –0.16 |
| Apr | 12.17 | 61.714 | 16 | 5 | 4.07 | –0.13 | –0.12 | –0.17 | –0.14 |
| Apr | 12.18 | 61.725 | 6 | 7 | 4.11 | –0.12 | –0.11 | –0.18 | –0.13 |
| Apr | 13.04 | 62.578 | 6 | 6 | 4.14 | 0.19 | 0.23 | 0.14 | 0.07 |
| Apr | 13.06 | 62.605 | 16 | 8 | 4.07 | 0.21 | 0.24 | 0.18 | 0.06 |
| Apr | 13.13 | 62.666 | 6 | 6 | 4.07 | 0.22 | 0.25 | 0.17 | 0.08 |
| Apr | 13.16 | 62.697 | 16 | 9 | 4.07 | 0.24 | 0.27 | 0.22 | 0.08 |
| Apr | 13.20 | 62.739 | 6 | 10 | 4.06 | 0.24 | 0.27 | 0.20 | 0.10 |
| Apr | 13.23 | 62.773 | 16 | 6 | 4.07 | 0.26 | 0.28 | 0.24 | 0.10 |
| Apr | 13.28 | 62.816 | 6 | 6 | 4.08 | 0.25 | 0.28 | 0.21 | 0.11 |
| Apr | 13.30 | 62.842 | 16 | 10 | 4.07 | 0.28 | 0.29 | 0.25 | 0.10 |
| Apr | 13.36 | 62.904 | 6 | 9 | 4.12 | 0.26 | 0.29 | 0.22 | 0.11 |
| Apr | 13.39 | 62.931 | 16 | 2 | 4.07 | 0.28 | 0.28 | 0.26 | 0.11 |
| Apr | 14.03 | 63.570 | 6 | 6 | 4.18 | 0.14 | 0.10 | 0.13 | 0.07 |
| Apr | 14.06 | 63.601 | 16 | 11 | 4.09 | 0.13 | 0.07 | 0.16 | 0.05 |
| Apr | 14.12 | 63.664 | 16 | 5 | 4.09 | 0.10 | 0.04 | 0.13 | 0.02 |
| Apr | 14.13 | 63.667 | 6 | 6 | 4.09 | 0.09 | 0.04 | 0.09 | 0.02 |
| Apr | 14.20 | 63.746 | 6 | 8 | 4.09 | 0.06 | 0.00 | 0.06 | 0.01 |
| Apr | 14.22 | 63.761 | 16 | 10 | 4.09 | 0.06 | –0.01 | 0.09 | 0.01 |
| Apr | 14.29 | 63.834 | 6 | 6 | 4.09 | 0.02 | –0.05 | 0.01 | –0.01 |
| Apr | 14.31 | 63.853 | 16 | 10 | 4.09 | 0.02 | –0.06 | 0.06 | –0.02 |
| Apr | 14.36 | 63.904 | 6 | 4 | 4.11 | 0.01 | –0.08 | 0.01 | –0.01 |
| Apr | 14.38 | 63.917 | 16 | 5 | 4.09 | 0.00 | –0.10 | 0.03 | –0.03 |
| Apr | 15.00 | 64.545 | 6 | 4 | 4.15 | –0.09 | –0.18 | –0.14 | –0.08 |
| Apr | 15.09 | 64.632 | 6 | 7 | 4.10 | –0.05 | –0.12 | –0.11 | –0.09 |
| Apr | 15.16 | 64.701 | 6 | 5 | 4.17 | –0.01 | –0.05 | –0.08 | –0.06 |
| Apr | 15.24 | 64.779 | 6 | 6 | 4.06 | 0.04 | 0.02 | –0.05 | –0.05 |
| Apr | 15.32 | 64.865 | 6 | 7 | 4.13 | 0.08 | 0.07 | 0.00 | –0.03 |
| Apr | 15.40 | 64.937 | 6 | 2 | 4.09 | 0.11 | 0.12 | 0.04 | –0.01 |
| Apr | 16.00 | 65.544 | 6 | 4 | 4.16 | 0.18 | 0.20 | 0.18 | 0.07 |
| Apr | 16.08 | 65.620 | 6 | 7 | 4.10 | 0.16 | 0.18 | 0.16 | 0.07 |
| Apr | 16.13 | 65.674 | 20 | 6 | 4.15 | 0.11 | 0.19 | 0.24 | 0.08 |
| Apr | 16.17 | 65.711 | 6 | 5 | 4.16 | 0.15 | 0.17 | 0.16 | 0.05 |
| Apr | 16.22 | 65.763 | 20 | 3 | 4.20 | 0.10 | 0.16 | 0.24 | 0.08 |
| Apr | 16.29 | 65.830 | 20 | 2 | 4.20 | 0.09 | 0.14 | 0.22 | 0.08 |
| Apr | 16.75 | 66.289 | 27 | 1 | 3.87 | –0.06 | –0.07 | 0.05 | –0.01 |
| Apr | 17.00 | 66.542 | 6 | 4 | 4.18 | 0.05 | 0.01 | 0.02 | 0.00 |
| Apr | 17.07 | 66.614 | 6 | 8 | 4.12 | 0.03 | 0.00 | 0.01 | –0.01 |
| Apr | 17.15 | 66.692 | 6 | 7 | 4.12 | 0.03 | 0.00 | 0.01 | –0.02 |
| Apr | 17.25 | 66.791 | 6 | 7 | 4.12 | 0.02 | –0.01 | –0.01 | –0.03 |
| Apr | 17.35 | 66.889 | 6 | 6 | 4.19 | 0.03 | 0.00 | 0.00 | –0.03 |
| Apr | 17.52 | 67.062 | 17 | 2 | 3.97 | 0.06 | 0.01 | 0.00 | –0.01 |
| Apr | 18.03 | 67.567 | 6 | 4 | 4.18 | –0.03 | –0.08 | –0.07 | –0.06 |
| Apr | 18.10 | 67.645 | 6 | 7 | 4.13 | –0.06 | –0.13 | –0.10 | –0.07 |
| Apr | 18.17 | 67.716 | 6 | 3 | 4.20 | –0.07 | –0.11 | –0.09 | –0.08 |
| Apr | 18.22 | 67.754 | 20 | 5 | 4.17 | –0.14 | –0.14 | –0.02 | –0.06 |
| Apr | 18.26 | 67.800 | 6 | 6 | 4.18 | –0.09 | –0.16 | –0.12 | –0.08 |
| Apr | 18.35 | 67.888 | 6 | 2 | 4.13 | –0.12 | –0.21 | –0.14 | –0.09 |
| Apr | 18.45 | 67.995 | 17 | 2 | 3.99 | –0.21 | –0.28 | –0.21 | –0.11 |
| Apr | 20.55 | 70.089 | 9 | 1 | 4.47 | 0.19 | 0.30 | 0.19 | 0.07 |
| Apr | 20.65 | 70.188 | 9 | 4 | 4.35 | 0.20 | 0.28 | 0.20 | 0.09 |
| Apr | 21.40 | 70.938 | 10 | 26 | 4.20 | 0.06 | 0.10 | 0.04 | 0.05 |
| Apr | 27.35 | 76.893 | 17 | 1 | 4.14 | 0.10 | 0.14 | 0.05 | 0.01 |
| Apr | 28.38 | 77.919 | 17 | 1 | 4.16 | 0.10 | –0.01 | 0.07 | 0.09 |
| Apr | 29.03 | 78.574 | 20 | 7 | 4.43 | –0.09 | –0.05 | 0.07 | 0.03 |
| Apr | 29.30 | 78.839 | 17 | 1 | 4.17 | –0.13 | –0.23 | –0.08 | –0.07 |
| Apr | 29.71 | 79.250 | 27 | 2 | 4.09 | –0.25 | –0.28 | –0.18 | –0.08 |
| Apr | 30.04 | 79.580 | 20 | 9 | 4.44 | –0.14 | –0.09 | –0.05 | –0.01 |
| Apr | 30.72 | 80.266 | 27 | 2 | 4.11 | –0.03 | –0.02 | 0.01 | 0.00 |
| May | 1.05 | 80.588 | 20 | 2 | 4.46 | –0.04 | 0.04 | 0.05 | 0.02 |
| May | 1.12 | 80.661 | 20 | 5 | 4.39 | –0.05 | 0.03 | 0.02 | 0.01 |
| May | 1.15 | 80.691 | 29 | 2 | 4.20 | –0.07 | –0.08 | –0.10 | 0.01 |
| May | 1.23 | 80.770 | 29 | 2 | 4.20 | –0.09 | –0.16 | –0.04 | 0.00 |
| May | 1.35 | 80.895 | 17 | 2 | 4.20 | –0.03 | 0.03 | 0.04 | –0.01 |
| May | 2.01 | 81.554 | 20 | 4 | 4.47 | –0.01 | 0.08 | 0.05 | 0.02 |
| May | 2.03 | 81.571 | 28 | 5 | 4.06 | 0.02 | –0.03 | 0.00 | 0.02 |
| May | 2.07 | 81.613 | 20 | 4 | 4.47 | –0.03 | 0.08 | 0.03 | 0.01 |
| May | 2.13 | 81.671 | 28 | 6 | 4.04 | 0.05 | 0.03 | 0.02 | 0.05 |
| May | 2.93 | 82.475 | 22 | 3 | 4.63 | 0.05 | 0.16 | 0.11 | 0.01 |
| May | 3.12 | 82.662 | 28 | 7 | 4.04 | –0.16 | –0.20 | –0.06 | 0.00 |
| May | 3.39 | 82.929 | 17 | 2 | 4.23 | –0.13 | –0.15 | –0.09 | –0.02 |
| May | 4.31 | 83.848 | 17 | 2 | 4.25 | 0.00 | 0.02 | –0.07 | –0.02 |
| May | 5.31 | 84.847 | 17 | 2 | 4.26 | 0.12 | 0.19 | 0.09 | 0.06 |
| May | 5.72 | 85.259 | 27 | 2 | 4.19 | –0.03 | 0.05 | 0.10 | 0.06 |
| May | 5.92 | 85.458 | 22 | 9 | 4.56 | 0.03 | 0.11 | 0.12 | 0.02 |
| May | 6.00 | 85.545 | 6 | 4 | 4.49 | 0.07 | 0.07 | 0.08 | 0.03 |
| May | 6.06 | 85.604 | 6 | 2 | 4.31 | 0.03 | 0.02 | 0.04 | 0.02 |
| May | 6.13 | 85.675 | 28 | 2 | 4.27 | –0.09 | –0.14 | 0.03 | 0.02 |
| May | 6.17 | 85.708 | 6 | 3 | 4.50 | 0.04 | 0.03 | 0.05 | 0.02 |
| May | 6.93 | 86.468 | 21 | 7 | 4.17 | –0.18 | –0.20 | –0.15 | –0.07 |
| May | 7.01 | 86.546 | 21 | 3 | 4.17 | –0.17 | –0.21 | –0.16 | –0.08 |
| May | 7.87 | 87.412 | 22 | 7 | 4.48 | –0.14 | –0.07 | –0.14 | –0.08 |
| May | 7.99 | 87.532 | 21 | 16 | 4.55 | –0.02 | 0.02 | –0.06 | –0.03 |
| May | 8.03 | 87.566 | 28 | 5 | 4.15 | –0.12 | –0.16 | –0.11 | –0.08 |
| May | 8.03 | 87.574 | 6 | 6 | 4.52 | 0.01 | –0.02 | –0.05 | –0.02 |
| May | 8.09 | 87.630 | 28 | 5 | 4.15 | –0.12 | –0.17 | –0.11 | –0.08 |
| May | 8.12 | 87.667 | 16 | 7 | 4.46 | 0.00 | –0.04 | –0.04 | –0.04 |
| May | 8.17 | 87.713 | 6 | 6 | 4.43 | 0.00 | –0.03 | –0.07 | –0.04 |
| May | 8.21 | 87.749 | 16 | 3 | 4.46 | 0.00 | –0.03 | –0.03 | –0.05 |
| May | 8.49 | 88.031 | 9 | 6 | 4.60 | 0.00 | 0.06 | –0.01 | 0.03 |
| May | 8.57 | 88.114 | 9 | 7 | 4.57 | –0.01 | 0.02 | –0.04 | 0.00 |
| May | 8.68 | 88.218 | 9 | 5 | 4.61 | –0.01 | 0.05 | –0.04 | –0.01 |
| May | 8.86 | 88.401 | 22 | 2 | 4.72 | –0.01 | 0.10 | 0.01 | 0.01 |
| May | 9.07 | 88.613 | 28 | 4 | 4.19 | 0.00 | –0.03 | –0.03 | 0.03 |
| May | 9.13 | 88.667 | 28 | 3 | 4.01 | –0.01 | –0.03 | –0.04 | 0.05 |
| May | 9.91 | 89.449 | 22 | 6 | 4.46 | 0.09 | 0.17 | 0.08 | 0.05 |
| May | 9.93 | 89.467 | 21 | 5 | 4.21 | 0.09 | 0.15 | 0.05 | 0.04 |
| May | 9.99 | 89.530 | 21 | 4 | 4.21 | 0.09 | 0.14 | 0.05 | 0.03 |
| May | 10.06 | 89.605 | 16 | 4 | 4.49 | 0.17 | 0.19 | 0.13 | 0.05 |
| May | 10.13 | 89.667 | 6 | 4 | 4.54 | 0.16 | 0.19 | 0.11 | 0.07 |
| May | 10.17 | 89.715 | 16 | 8 | 4.49 | 0.15 | 0.16 | 0.12 | 0.05 |
| May | 10.19 | 89.727 | 6 | 3 | 4.51 | 0.16 | 0.18 | 0.11 | 0.07 |
| May | 11.02 | 90.561 | 6 | 3 | 4.57 | –0.05 | –0.12 | –0.05 | 0.00 |
| May | 11.15 | 90.688 | 6 | 5 | 4.47 | –0.09 | –0.17 | –0.10 | –0.01 |
| May | 11.93 | 91.473 | 21 | 6 | 4.26 | 0.12 | 0.18 | 0.02 | –0.01 |
| May | 12.92 | 92.459 | 21 | 4 | 4.24 | 0.05 | 0.11 | 0.06 | 0.02 |
| May | 13.73 | 93.270 | 27 | 1 | 4.29 | –0.21 | –0.21 | –0.04 | –0.04 |
| May | 13.97 | 93.511 | 21 | 3 | 4.26 | –0.16 | –0.18 | –0.12 | –0.05 |
| May | 14.87 | 94.408 | 22 | 5 | 4.65 | –0.22 | –0.13 | –0.12 | –0.09 |
| May | 15.86 | 95.405 | 22 | 2 | 4.80 | –0.10 | –0.04 | –0.09 | –0.02 |
| May | 19.70 | 99.239 | 27 | 2 | 4.35 | 0.13 | 0.20 | 0.20 | 0.13 |
| May | 21.63 | 101.167 | 9 | 3 | 4.91 | 0.08 | 0.08 | 0.10 | 0.08 |
| May 22.48 | 102.021 | 9 | 2 | 4.92 | –0.02 | –0.05 | 0.01 | 0.06 |
| May 26.30 | 105.837 | 17 | 1 | 4.50 | 0.03 | 0.08 | 0.00 | 0.02 |
| May 27.33 | 106.871 | 17 | 2 | 4.51 | 0.08 | 0.14 | 0.08 | 0.02 |
| May 28.30 | 107.840 | 17 | 1 | 4.52 | –0.03 | –0.01 | 0.01 | 0.00 |
| May 29.29 | 108.833 | 17 | 1 | 4.53 | –0.16 | –0.19 | –0.11 | 0.02 |
| May 30.29 | 109.835 | 17 | 1 | 4.54 | –0.12 | –0.15 | –0.14 | –0.02 |
| May 30.99 | 110.533 | 6 | 2 | 4.58 | 0.09 | 0.10 | –0.04 | 0.04 |
| May 31.06 | 110.605 | 6 | 3 | 4.75 | 0.08 | 0.10 | –0.03 | 0.02 |
| May 31.30 | 110.837 | 17 | 1 | 4.55 | 0.14 | 0.21 | 0.05 | 0.04 |
| May 31.72 | 111.260 | 27 | 2 | 4.46 | 0.09 | 0.19 | 0.15 | 0.09 |
| Jun 1.01 | 111.552 | 6 | 3 | 4.57 | 0.16 | 0.16 | 0.09 | 0.07 |
| Jun 2.03 | 112.572 | 6 | 4 | 4.77 | 0.02 | –0.04 | 0.02 | 0.01 |
| Jun 2.29 | 112.836 | 17 | 1 | 4.56 | –0.10 | –0.10 | –0.05 | 0.00 |
| Jun 3.01 | 113.552 | 6 | 3 | 4.85 | 0.09 | 0.11 | 0.04 | 0.02 |
| Jun 3.09 | 113.634 | 6 | 2 | 4.65 | 0.12 | 0.12 | 0.04 | 0.02 |
| Jun 3.29 | 113.834 | 17 | 1 | 4.57 | 0.10 | 0.13 | 0.07 | –0.01 |
| Jun 3.47 | 114.015 | 9 | 2 | 5.02 | 0.08 | 0.15 | 0.08 | –0.01 |
| Jun 3.53 | 114.076 | 9 | 4 | 4.90 | 0.08 | 0.18 | 0.07 | –0.01 |
| Jun 4.01 | 114.556 | 6 | 4 | 4.78 | 0.08 | 0.06 | 0.05 | 0.01 |
| Jun 4.29 | 114.831 | 17 | 1 | 4.58 | 0.01 | 0.03 | 0.02 | –0.04 |
| Jun 5.71 | 116.251 | 27 | 1 | 4.50 | –0.17 | –0.17 | –0.05 | 0.00 |
| Jun 6.02 | 116.560 | 6 | 3 | 4.82 | –0.05 | –0.11 | –0.01 | –0.09 |
| Jun 6.09 | 116.631 | 6 | 1 | 4.74 | –0.05 | –0.16 | –0.05 | –0.03 |
| Jun 6.33 | 116.870 | 17 | 1 | 4.59 | –0.16 | –0.20 | –0.12 | –0.06 |
| Jun 7.01 | 117.555 | 6 | 3 | 4.88 | 0.00 | –0.05 | –0.05 | –0.01 |
| Jun 7.28 | 117.818 | 17 | 1 | 4.60 | 0.05 | 0.08 | –0.03 | –0.01 |
| Jun 8.34 | 118.877 | 17 | 1 | 4.60 | 0.15 | 0.24 | 0.10 | 0.13 |
| Jun 9.32 | 119.856 | 17 | 1 | 4.61 | –0.01 | 0.00 | 0.05 | 0.10 |
| Jun 10.30 | 120.845 | 17 | 1 | 4.62 | 0.03 | 0.04 | 0.02 | 0.09 |
| Jun 13.99 | 124.530 | 28 | 2 | 4.64 | –0.22 | –0.30 | –0.18 | –0.05 |
| Jun 25.01 | 135.552 | 6 | 2 | 4.86 | –0.06 | –0.23 | –0.10 | –0.05 |
| Jun 26.01 | 136.548 | 6 | 2 | 4.86 | –0.03 | –0.11 | –0.06 | –0.07 |
| Jun 26.98 | 137.518 | 28 | 1 | 4.89 | –0.04 | 0.01 | 0.00 | –0.04 |
| Jun 30.98 | 141.519 | 28 | 1 | 4.91 | –0.07 | –0.11 | 0.04 | –0.03 |
| Jul 5.99 | 146.534 | 28 | 2 | 4.75 | 0.13 | –0.03 | 0.01 | –0.06 |
| Jul 6.98 | 147.516 | 28 | 2 | 4.93 | 0.00 | 0.04 | 0.07 | –0.03 |

[a] Time from perihelion (1986 Feb 9.459)
[b] Telescope ID from Table 1
[c] Number of observations within each ~2 hr bin which were averaged together
[d] Log of the mean projected aperture radius in km
[e] Residual production rates after trends for $r_H$-dependence were removed, in log $Q$ (molecules $s^{-1}$) for the gases CN, $C_3$, and $C_2$, and log $A(0°)f\rho$ (cm) for the green continuum at 4845 Å

**Table 4**
Apparent Rotation Periods as a Function of Time

| $\Delta T$ Mid-point[a] (day) | Apparent Rotation Periods: CN Lightcurve | CN PDM | $C_3$ Lightcurve | $C_3$ PDM | $C_2$ Lightcurve | $C_2$ PDM | Green Continuum Lightcurve | Green Continuum PDM |
|---|---|---|---|---|---|---|---|---|
| Pre-perihelion | | | | | | | | |
| –94.5 | 7.49±.05 | 7.69 | 7.57±.15 | — | 7.47±.05 | 7.69 | — | — |
| –87.5 | 7.57±.10 | 7.41 | 7.56±.10 | 7.69 | 7.56±.10 | 8.00 | — | — |
| –80.5 | 7.52±.10 | 7.46 | 7.51±.10 | 7.46 | 7.55±.10 | 7.46 | — | — |
| –73.5 | 7.47±.10 | 7.35 | 7.50±.15 | 7.58 | 7.54±.08 | 7.30 | — | — |
| –66.5 | 7.59±.15 | 7.58 | 7.39±.10 | 7.41 | 7.59±.10 | 7.69 | — | — |
| –59.5 | 7.49±.10 | 7.69 | 7.41±.15 | 7.69 | 7.44±.10 | 7.69 | — | — |
| –52.5 | 7.47±.10 | 7.58 | 7.43±.15 | 7.58 | 7.44±.10 | 7.87 | — | — |
| –45.5 | 7.47±.10 | 8.13 | 7.48±.20 | 7.69 | 7.40±.15 | 7.94 | — | — |
| –38.5 | 7.52±.10 | — | 7.44±.10 | 8.26 | 7.40±.10 | 7.35 | — | — |
| Post-perihelion | | | | | | | | |
| +24.5 | 7.56±.05 | 7.46 | 7.50±.02 | 7.52 | 7.52±.03 | 7.46 | 7.52±.05 | 7.58 |
| +31.5 | 7.57±.03 | 7.58 | 7.55±.02 | 7.58 | 7.50±.05 | 7.58 | 7.51±.05 | 7.58 |
| +38.5 | 7.44±.03 | 7.30 | 7.45±.03 | 7.30 | 7.39±.03 | 7.30 | 7.41±.05 | 7.30 |
| +45.5 | 7.35±.03 | 7.14 | 7.35±.01 | 7.14 | 7.36±.03 | 7.46 | 7.40±.05 | 7.46 |
| +52.5 | 7.33±.02 | 7.14 | 7.34±.01 | 7.35 | 7.35±.03 | 7.41 | 7.31±.03 | 7.41 |
| +59.5 | 7.35±.02 | 7.41 | 7.34±.02 | 7.41 | 7.37±.03 | 7.41 | 7.32±.03 | 7.30 |
| +66.5 | 7.35±.02 | 7.35 | 7.34±.02 | 7.35 | 7.34±.02 | 7.35 | 7.32±.03 | 7.35 |
| +73.5 | 7.34±.02 | — | 7.35±.03 | 7.46 | 7.37±.02 | — | 7.35±.03 | — |
| +80.5 | 7.33±.02 | 7.30 | 7.36±.05 | 7.30 | 7.32±.03 | 7.30 | 7.38±.03 | 7.41 |
| +87.5 | 7.32±.02 | 7.35 | 7.35±.03 | 7.35 | 7.29±.03 | 7.30 | 7.33±.05 | 7.35 |
| +94.5 | 7.25±.05 | 7.09 | 7.28±.05 | 7.09 | 7.30±.05 | 7.09 | 7.32±.05 | 8.13 |
| +101.5 | 7.25±.05 | 7.09 | 7.28±.08 | 7.09 | 7.24±.03 | 7.09 | 7.23±.05 | 7.04 |
| +108.5 | 7.27±.03 | 7.09 | 7.28±.08 | 7.09 | 7.24±.03 | 7.14 | 7.25±.05 | 7.09 |
| +115.5 | 7.21±.05 | 7.41 | 7.28±.05 | 7.41 | 7.28±.05 | 7.41 | 7.24±.05 | 7.81 |

[a]Mid-point of the three-week interval used for each period determination